\documentclass[lettersize,journal]{IEEEtran}
\IEEEoverridecommandlockouts
\usepackage{cite}
\usepackage{amsmath,amssymb,amsfonts}
\usepackage{algorithmic}
\usepackage{graphicx}
\usepackage{textcomp}
\usepackage{xcolor}
\usepackage{array}
\usepackage{multirow}
\usepackage{makecell}
\usepackage{booktabs}
\usepackage{subfigure}

\usepackage[pagebackref=false,breaklinks=false,letterpaper=true,colorlinks=true,urlcolor=red,citecolor=blue,linkcolor=blue, anchorcolor=blue, bookmarks=true]{hyperref}
\def\BibTeX{{\rm B\kern-.05em{\sc i\kern-.025em b}\kern-.08em
    T\kern-.1667em\lower.7ex\hbox{E}\kern-.125emX}}
\begin{document}

\title{FSOS-AMC: Few-Shot Open-Set Learning for Automatic Modulation Classification Over Multipath Fading Channels\\
\thanks{

This work was supported in part by the National Natural Science Foundation of China under Grant 62222107 and the Yangtze River Delta Science and Technology Innovation Community Joint Research (Basic Research) Project under Grant BK20244006. 
The work was presented in part to the 2024 16th International Conference on Wireless Communications and Signal Processing (WCSP), Hefei, China, 2024 \cite{zhang2024few}. (\emph{Corresponding author: Fuhui Zhou})

H. Zhang and Q. Wu are with the College of Electronic and Information Engineering, F. Zhou is with the College of Artificial Intelligence, Nanjing University of Aeronautics and Astronautics, Nanjing 211106 China. They are also with the Key Laboratory of Dynamic Cognitive System of Electromagnetic Spectrum Space (Nanjing University of Aeronautics and Astronautics) and with the Ministry of Industry and Information Technology, Nanjing, 211106, China (emails: haozhangcn@nuaa.edu.cn, zhoufuhui@ieee.org, wuquhui2014@sina.com)

C. Yuen is with the School of Electrical and Electronic Engineering, Nanyang Technological University, Singapore 639798 (email: chau.yuen@ntu.edu.sg). 

}
}

\author{

Hao Zhang, \IEEEmembership{Graduate Student Member, IEEE}, Fuhui Zhou, \IEEEmembership{Senior Member, IEEE}, 
\\Qihui Wu, \IEEEmembership{Fellow, IEEE}, 
and Chau Yuen, \IEEEmembership{Fellow, IEEE}
}

\maketitle

\begin{abstract}
Automatic modulation classification (AMC) plays a vital role in advancing future wireless communication networks. Although deep learning (DL)-based AMC frameworks have demonstrated remarkable classification capabilities, they typically require large-scale training datasets and assume consistent class distributions between training and testing data-prerequisites that prove challenging in few-shot and open-set scenarios. To address these limitations, we propose a novel few-shot open-set automatic modulation classification (FSOS-AMC) framework that integrates a multi-sequence multi-scale attention network (MS-MSANet), meta-prototype training, and a modular open-set classifier. The MS-MSANet extracts features from multi-sequence input signals, while meta-prototype training optimizes both the feature extractor and the modular open-set classifier, which can effectively categorize testing data into known modulation types or identify potential unknown modulations. Extensive simulation results demonstrate that our FSOS-AMC framework achieves superior performance in few-shot open-set scenarios compared to state-of-the-art methods. Specifically, the framework exhibits higher classification accuracy for both known and unknown modulations, as validated by improved accuracy and area under the receiver operating characteristic curve (AUROC) metrics. Moreover, the proposed framework demonstrates remarkable robustness under challenging low signal-to-noise ratio (SNR) conditions, significantly outperforming existing approaches.
\end{abstract}


\section{Introduction}
\IEEEPARstart{A}{UTOMATIC} modulation classification (AMC) is crucial for future wireless communication systems because it facilitates a range of applications including cognitive radio and intelligent spectrum management. AMC is the process of identifying the modulation scheme of a received signal, which is essential for both civilian and military applications. In civilian applications, AMC can be used for identifying unauthorized signals by detecting the modulation schemes, while in military applications, AMC facilitates abnormal signal detection and interference detection. With the further development of next-generation wireless communication systems, such as sixth-generation (6G), the AMC problem becomes more challenging due to the increasing number of modulation schemes and the dynamic wireless environment. Thus, it is essential to develop efficient and robust AMC frameworks. 

AMC aims at classifying the received signals into one of the known modulation schemes in non-cooperative communication environments without prior knowledge of the transmission parameters. AMC has numerous applications across the military and civilian domains, including decoding of intercepted transmissions, adaptive demodulation, and dynamic spectrum access (DSA). With the deep integration of artificial intelligence (AI) and future wireless communication, AMC has attracted extensive attention in the research community \cite{zhang2024few}. 

\subsection{Related Works and Motivations}

\begin{figure*}
	\centering
	\includegraphics[width=0.99\linewidth]{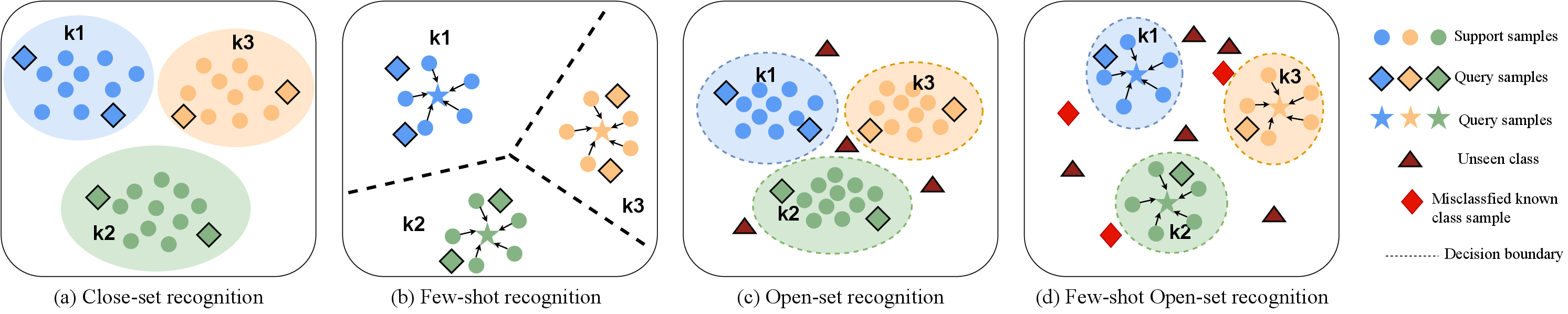} 
	\caption{Illustration of the \emph{few-shot open-set} recognition compared to (a) closed-set recognition, (b) few-shot recognition, and (c) open-set recognition.}
	\label{fig:fsos}
\end{figure*}

Traditional AMC schemes include likelihood-based (LB) schemes and feature-based (FB) schemes \cite{zhang2021novel,zhang2024sswsrnet}. 
The LB schemes formulate AMC as a hypothesis testing problem, which requires pre-known signal-to-noise ratio (SNR) and frequency offsets \cite{hameed2009likelihood,xu2010likelihood}. 
The average likelihood ratio test (ALRT), the general likelihood ratio test (GLRT), hybrid likelihood ratio test (HLRT) methods are three main kinds of LB approaches. 
Although LB approaches, in theory, guarantee optimal classification results from the Bayesian sense, they have high computational complexity and require prior knowledge of channel parameters, which is not applicable in practice. 
FB methods have been proven to provide suboptimal solutions more efficiently due to their lower computational complexity \cite{hazza2013overview}. The type of modulation is identified by extracting relevant features from the received signals. High-quality features are crucial for achieving robust performance with minimal complexity. Previous research has explored various features for AMC, including wavelet transforms \cite{boutte2009hybrid}, cyclostationary features \cite{xie2017cyclic}, and higher-order cumulants \cite{han2016low}. In the classification phase, conventional classifiers such as random forests \cite{zhang2017method}, $k$-nearest neighbors (KNN) \cite{aslam2012automatic}, Naive Bayes \cite{li2015bayesian}, support vector machine (SVM) \cite{park2008automatic,zhang2017modulation}, and multi-layer perception (MLP) \cite{wong2004automatic} are used. However, these methods are often time-intensive as they rely on the extraction of handcrafted features, which necessitates substantial domain knowledge and engineering. 
However, the FB methods rely on the handcrafted features, which may not be the best representation for AMC. Moreover, the FB methods may not be suitable for the complex and dynamic wireless communication environment. 

In response to the rapid advancements in AI, deep learning (DL) has been demonstrated powerful in solving various applications including computer vision \cite{zhang2019recent,xu2022automatic}, geoscience \cite{xu2024soil,zhang2023geoscience}, as well as wireless communication \cite{letaief2019roadmap,zhang2021novel,hu2022robust}. 
DL-based AMC frameworks have been exploited, which can be capable of autonomously learning features directly from raw signals \cite{zhang2021automatic}. 
For example, the authors in \cite{o2016convolutional} first developed a RadioML dataset and proposed a convolutional neural network (CNN)-based AMC framework for solving the AMC problem,  achieving a high classification accuracy. 
Later, a large-scale dataset was proposed in \cite{o2018over} and a ResNet-based \cite{he2016deep} AMC framework was proposed for solving the AMC problem. 
However, the existing deep learning-based AMC methods relied on two dataset assumptions, large-scale training datasets and the same class pool between the training and testing data, \emph{i.e.} a closed-set scenario. 
These assumptions failed under two scenarios, 1) \emph{few-shot conditions}, where it is hard to collect enough labeled data and 2) \emph{open-set scenarios}, where the testing classes do not appear during training. 
While existing deep learning approaches have shown promise in modulation classification, their performance degrades significantly under realistic channel conditions, especially multipath fading. This challenge is particularly acute in few-shot and open-set scenarios where the model must generalize from limited training samples while dealing with channel distortions. The widely-used RadioML 2016.10A dataset, while valuable, inadequately represents authentic channel conditions, particularly the dynamic nature of propagation channels and severe multipath effects. This limitation has led to a gap between theoretical performance and practical deployability of existing AMC solutions.

\emph{Closed-set AMC frameworks:} 
As shown in Fig. \ref{fig:fsos} (a), the closed-set scenario is the most common scenario in AMC, where the training and testing data are from the same set of modulation schemes. 
Various deep learning-based AMC frameworks have been proposed, such as CNN-based \cite{zhang2021novel}, recurrent neural network (RNN)-based \cite{hu2019deep}, and hybrid frameworks. 
The works \cite{o2016convolutional} and \cite{o2018over} introduced the RadioML datasets, which are widely used for AMC research. 
Zhang \emph{et al.} \cite{zhang2021novel} proposed a novel multi-scale convolutional neural network-based AMC framework, which can achieve high classification accuracy. 
The authors in \cite{hu2019deep} exploited an RNN-based AMC framework, demonstrating that RNN-based classifier is robust to the uncertain noise conditions. 
To solve the AMC problem under few-sample conditions, the authors in \cite{zhang2024sswsrnet} proposed a novel few-shot AMC framework by using semi-supervised learning, which can achieve high classification accuracy with a few samples. 
In summary, while CNNs and RNNs have driven initial wireless deep learning research, a key limitation that remains is their reliance on large labeled datasets for training. 

\emph{Signal representations:} FB methods focus on extracting various features from the received signals. However, when DL methods are used, most of the existing works only use the IQ signal representation for several reasons. First, the IQ signal representation is the most common representation for AMC. Second, the IQ signal representation is easy to extract and can be directly fed into the DL model. However, the IQ signal representation may not be the best representation for AMC. The amplitude and phase (AP) representation and power spectral density (PSD) representation can provide more information about the signal, which may be useful for AMC. 
Qiu \emph{et al.} \cite{qiu2023deepsig} proposed DeepSIG, which is a hybrid heterogeneous modulation classification architecture. It combines RNN, CNN, and GNN models to process radio signals with multiple input formats, including IQ sequences, images mapped from IQ signals and graphs converted from IQ signals. The DeepSIG model can outperform single-input methods in classification accuracy, especially in few-shot scenarios. 
However, the combination of signal representations from images and graphs may not be the best representation for AMC for its complexity. 
Thus, it is essential to explore the best signal representation for AMC.

\emph{Channel effects}: Deep learning models for modulation classification must be robust to diverse channel conditions in real-world applications. While the widely-used RadioML2016.10a dataset \cite{o2016convolutional} has contributed significantly to the field, it inadequately represents authentic channel conditions, particularly the dynamic nature of propagation channels and severe multipath effects. Although the RadioML2018.01a dataset \cite{o2018over} attempted to address this limitation by incorporating over-the-air recordings of 24 digital and analog modulation types, its reliance on uncontrollable propagation environments makes it impossible to analyze the relationship between specific channel conditions and model performance. Additionally, being generated in a controlled laboratory environment, the dataset lacks diversity in crucial channel parameters such as fading and the number of taps. Building upon the foundation of RML2016.10a, the authors in \cite{sathyanarayanan2023rml22} introduced RML22, which improves upon the original dataset's methodology and data quality through a data-centric approach. Additionally, being generated in a controlled laboratory environment, the dataset lacks diversity in crucial channel parameters such as fading and the number of taps.
To address these channel-related challenges, Wu \emph{et al.} \cite{wu2008novel} developed a cumulant-based feature extraction method for AMC that achieves high classification accuracy under multipath fading channels. However, there remains a pressing need for datasets and models that can effectively handle diverse and realistic channel conditions.

\emph{Few-shot AMC frameworks:} Few-shot AMC is a scenario where the model is trained with a few samples of each modulation scheme. 
Recent works can be categorized into three categories, namely, few-sample-based methods, support-data-based methods, and synthetic-data-based methods \cite{zhang2024sswsrnet}. 
For few-sample-based methods, they only utilize the labeled data with few samples by using specifically designed architectures and transfer learning. Li \emph{et al.} \cite{li2020automatic} introduced an AMR-CapsNet for AMC that achieves high accuracy under limited training data. The authors in [28] proposed a transfer learning method for AMC. 
The authors in \cite{zhang2024sswsrnet} proposed a novel few-shot AMC framework by using semi-supervised learning, which can achieve high classification accuracy with a few samples. 
Existing few-shot AMC frameworks focused on the few-shot learning problem under the closed-set scenario, without considering the open-set recognition problem. 

\emph{Open-set AMC frameworks:} Open-set recognition (OSR) is a scenario where the model is trained with a set of modulation schemes, but the testing data may contain modulation schemes that are not seen during training. 
Bendale \emph{et al.} \cite{bendale2016towards} proposed the OpenMax model, replacing the softmax layer in the deep network with the OpenMax layer for OSR. 
Chen \emph{et al.} \cite{chen2023open} proposed a metric-based OSR framework, that can achieve high classification accuracy. 
The authors in \cite{chen2024boosting} exploited a feature space singularity-based framework, which can improve discrimination between in-distribution and out-of-distribution features by promoting compact classification boundaries and reducing feature overlap. 
However, the existing open-set AMC frameworks still rely on large-scale training data. 

As illustrated in Fig. \ref{fig:fsos} (d), few-shot open-set recognition represents a significantly more challenging paradigm compared to traditional closed-set, few-shot, and open-set recognition scenarios. This challenge becomes particularly relevant in automatic modulation classification (AMC) within complex and dynamic wireless communication environments, where systems must accurately classify signals using limited training samples while simultaneously identifying potential unknown modulation types. To address these practical challenges, we propose a novel few-shot open-set AMC framework that demonstrates robust classification performance for both known and unknown modulation types, even with limited training data.

\subsection{Contributions and Organization}

Motivated by the challenges of few-shot open-set AMC under practical multipath fading channels, we propose a novel few-shot open-set automatic modulation classification (FSOS-AMC) framework. The FSOS-AMC framework consists of three main parts designed to address both limited training samples and channel impairments: a channel-aware multi-scale attention network, the meta-prototype training, and a modular open-set classifier. The multi-scale attention network adaptively learns robust multi-scale features from signals experiencing multipath fading, the meta-prototype training is used to train the feature extractor to be channel-invariant in a few-shot learning manner, and the modular open-set classifier enables accurate classification of testing data into known or potential unknown modulations even under severe channel conditions. Simulation results demonstrate that the proposed FSOS-AMC framework can achieve high classification accuracy for both known and unknown modulations compared to state-of-the-art methods across various multipath fading scenarios. Moreover, the proposed FSOS-AMC framework maintains robust performance under both low SNR conditions and challenging multipath environments with minimal confusion between known and unknown modulations.

\begin{enumerate}
    \item A novel few-shot open-set automatic modulation classification (FSOS-AMC) framework is proposed to achieve robust classification for both known and unknown modulations under practical multipath fading channel conditions. The FSOS-AMC framework incorporates three key components: a channel-aware multi-scale attention network, meta-prototype training for few-shot learning, and a modular open-set classifier.
    
    \item A multi-sequence multi-scale attention network (MS-MSANet) is developed to extract channel-robust features from multi-sequence input signals affected by multipath fading. The meta-prototype training is designed to learn channel-invariant features with limited training samples, while the modular open-set classifier enables accurate classification of both known and potential unknown modulations even under severe channel impairments.
    
    \item Extensive simulation results under various multipath fading scenarios demonstrate the superiority of our proposed FSOS-AMC framework compared with other DL-based schemes in terms of classification accuracy for both known and unknown modulations. The framework maintains high performance even under challenging channel conditions. Moreover, the visualized features validate the framework's ability to learn discriminative and channel-robust representations, highlighting its effectiveness in practical wireless environments.
\end{enumerate}

The structure of this paper is organized as follows. 
Section \ref{sec:preliminaries} introduces the preliminaries, including the signal model and the problem formulation. 
Section \ref{sec:proposed_method} details our proposed FSOS-AMC framework, including the multi-sequence representation, the multi-sequence multi-scale attention network, meta-prototype training, and modular open-set classifier. 
Section \ref{sec:experiments_results} discusses the simulation results. The paper concludes in Section \ref{sec:conclusion}.

\section{Preliminaries}
\label{sec:preliminaries}

In this section, the signal model is briefly introduced, and the signal preprocessing is implemented to acquire a good signal representation. Then, the problem formulation is presented. 

\subsection{Signal Model}
Assume that the wireless communication system is equipped with a transmitter, a channel and a receiver. The transmitter modulates the information bits into a signal, which is transmitted through the channel and received by the receiver. The received signal $x(n)$ can be denoted as 
\begin{equation}
    x(n) = s(n)*h(n) + w(n), n = 0, 1, \cdots, N,
\end{equation}
where $s(n)$ is the transmitted signal, $h(n)$ is the channel impulse response, $w(n)$ is the additive white Gaussian noise (AWGN), and $N$ is the number of samples. 

The received signal $x(n)$ can be transferred into an in-phase and quadrature (I/Q) signal $\boldsymbol{x}_{\text{IQ}}(n)$ as a vector, given as
\begin{equation}
	\boldsymbol{x}_{\text{IQ}}(n) = \begin{bmatrix}
	\Re\{x(1), x(2), \cdots, x(N)\} \\
	\Im\{x(1), x(2), \cdots, x(N)\}
	\end{bmatrix},
\label{eq:iq_signal}
\end{equation}
where $\Re\{\cdot\}$ and $\Im\{\cdot\}$ denote the operators of the real and imaginary parts of the complex signal, respectively.

\subsection{Problem Formulation}
The AMC problem can be formulated as a multi-class classification problem. Given a set of $N$ modulation schemes, the goal is to classify the received signal into one of the $N$ classes. 
For closed-set AMC, the training data and testing data are from the same set of modulation schemes. 
It can be formulated as
\begin{equation}
	\hat{y} = \arg\max_{y} p(y|\boldsymbol{x}_{\text{IQ}}),
\end{equation}
where $\hat{y}$ is the predicted modulation scheme, $p(y|\boldsymbol{x}_{\text{IQ}})$ is the probability distribution of the modulation schemes given the received signal $\boldsymbol{x}_{\text{IQ}}$.

For open-set AMC, the model is trained with a set of modulation schemes, but the testing data may contain modulation schemes that are not seen during training.
Given a set of training data $D_{tr}=\{(x_1,y_1),(x_2,y_2),\cdots,(x_n,y_n)\}$ with $n$ labeled instances and $N$ known classes, $y_i\in{1,2,\cdots,N}$ denotes the label of the $i$-th instance. 
The goal of open-set AMC is to classify the testing data $D_{te}=\{x_1,x_2,\cdots,x_m\}$ with $m$ instances into one of the $N$ known classes or potential unknown classes, denoted as $N+1$. 
It should be noted that the unknowns may belong to various classes, and their specific classifications are not the primary focus of the open-set recognition task.

\section{Proposed FSOS-AMC Framework}
\label{sec:proposed_method}
In this section, we present the proposed few-shot open-set automatic modulation classification (FSOS-AMC) framework. Firstly, an overview of the proposed FSOS-AMC framework is presented. Then, the multi-sequence representation is introduced to acquire a good signal representation, followed by the multi-sequence multi-scale attention network (MS-MSANet) for feature extraction. Additionally, the meta-prototype training is adopted to train the feature extractor in a few-shot learning manner. Finally, the modular open-set classifier is used to classify the testing data into one of the known or potential unknown modulations.

\subsection{Overview of the Proposed Few-Shot Open-Set AMC (FSOS-AMC) Framework}
\begin{figure*}
	\centering
	\includegraphics[width=0.95\linewidth]{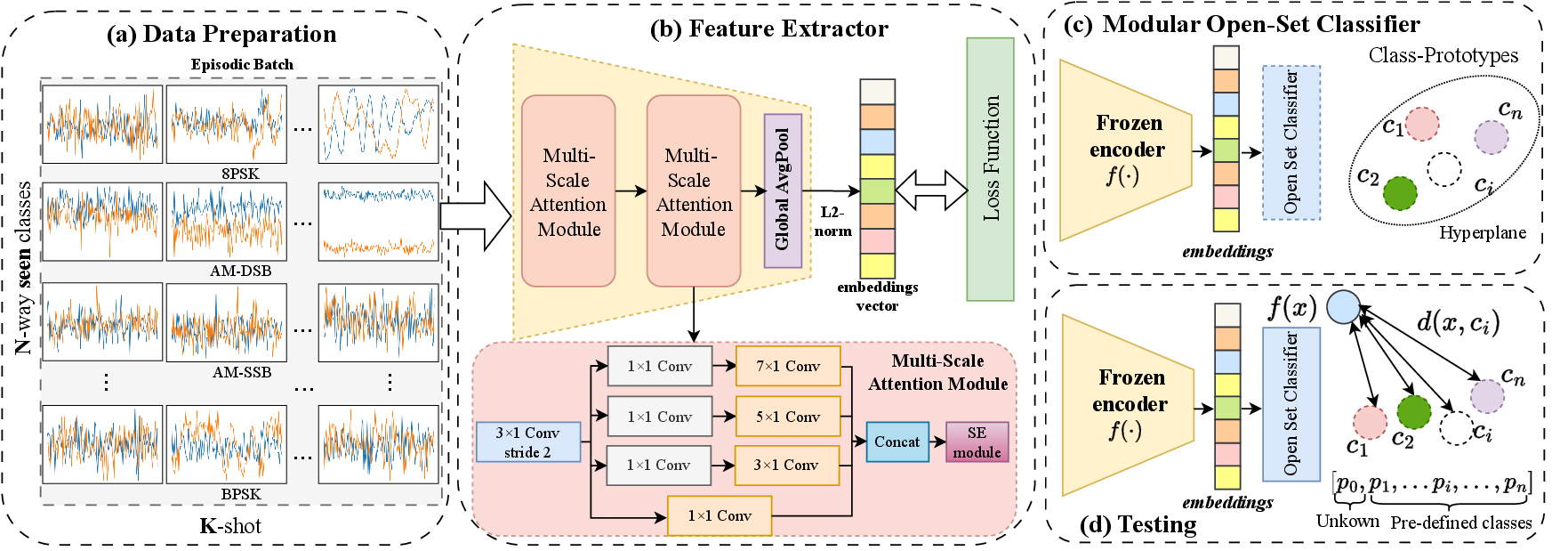} 
	\caption{Overview of the proposed few-shot open-set automatic modulation classification (FSOS-AMC) framework.}
	\label{fig:framework}
\end{figure*}

The proposed few-shot open-set automatic modulation classification (FSOS-AMC) framework is shown in Fig. \ref{fig:framework}. 
The FSOS-AMC framework consists of three main parts, multi-scale attention feature extractor, meta-prototype training, and modular open-set classifier. 
The multi-scale attention feature extractor is used to extract the features from the input signal by using a combination of a multi-scale module and a channel attention module. 
The meta-prototype training can train the feature extractor in a few-shot learning manner, which can learn the features from a small number of samples. 
Finally, the modular open-set classifier can classify the testing data into one of the known or potential unknown modulations. 

\subsection{Multi-Sequence Representations}
\begin{figure*}
\centering
\subfigure[]{\includegraphics[height=3.5cm]{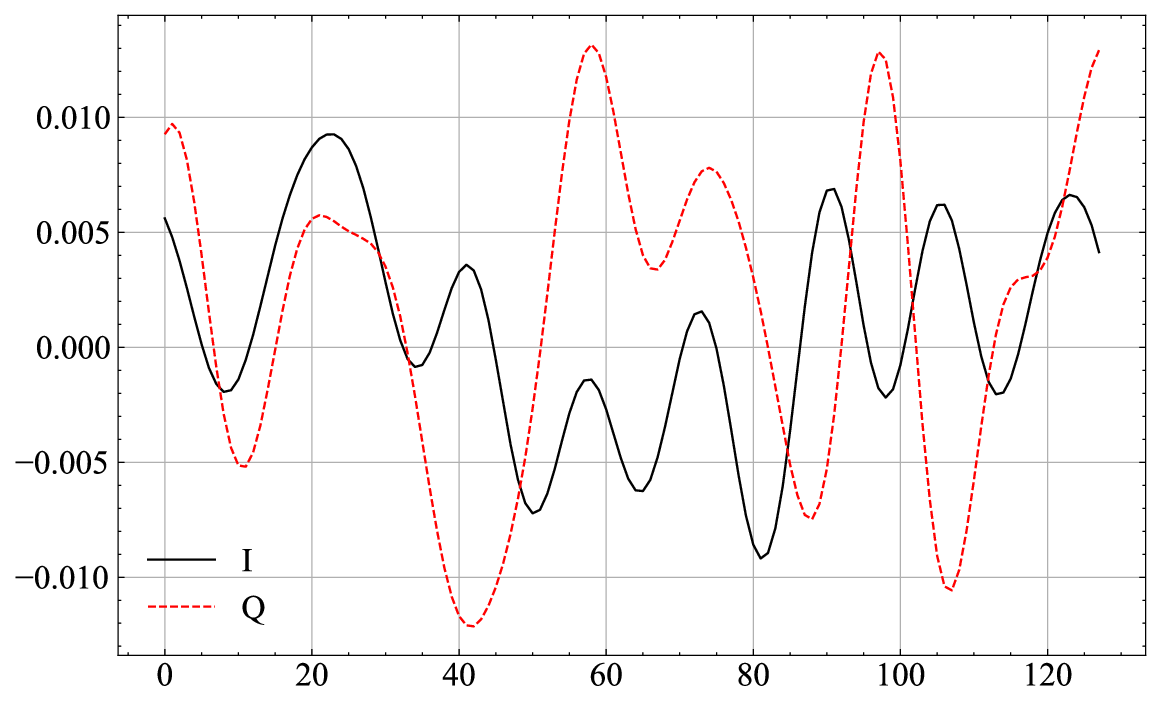}}
\subfigure[]{\includegraphics[height=3.5cm]{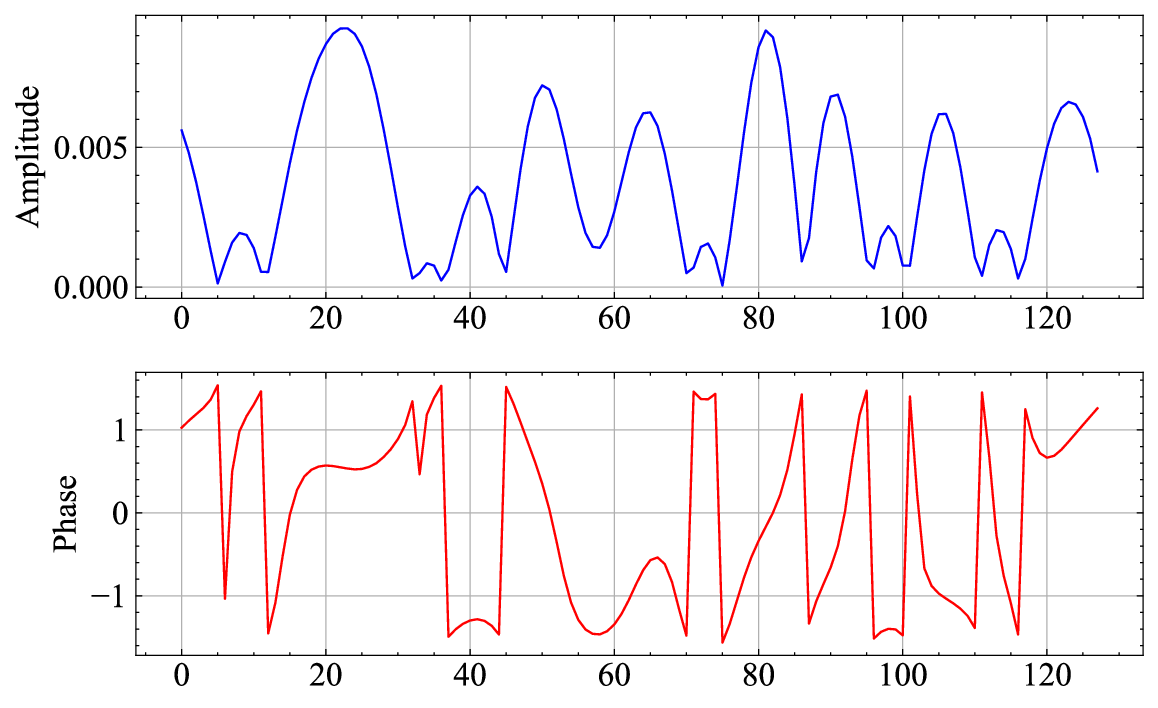}}
\subfigure[]{\includegraphics[height=3.5cm]{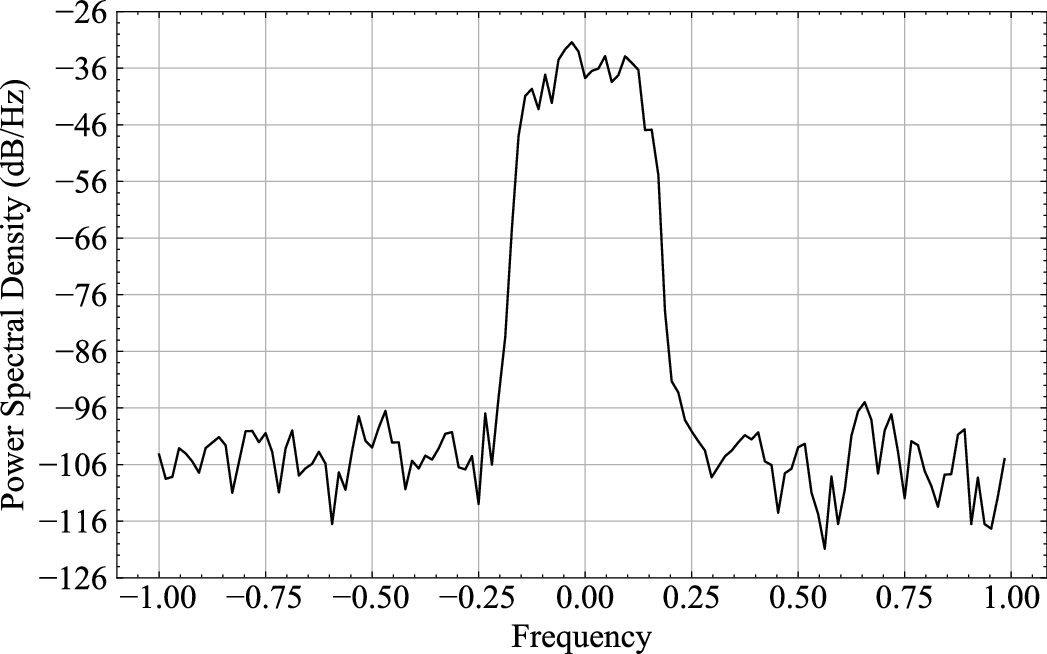}}
\DeclareGraphicsExtensions.
\caption{Different representations of the received signal of RadioML 2016.10A \cite{o2016convolutional}, including (a) IQ signal, (b) amplitude and phase (AP), and (c) power spectral density (PSD).}
\label{fig:data_representation2016}
\end{figure*}

\begin{figure*}
\centering
\subfigure[]{\includegraphics[height=3.5cm]{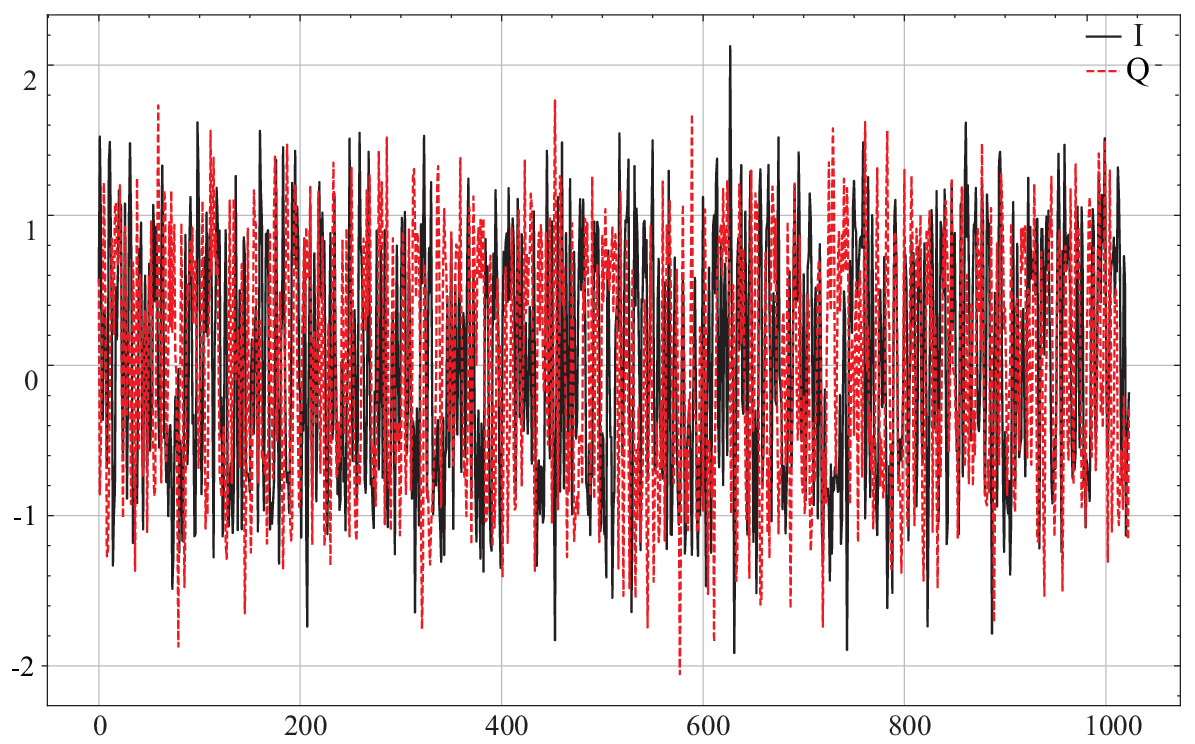}}
\subfigure[]{\includegraphics[height=3.5cm]{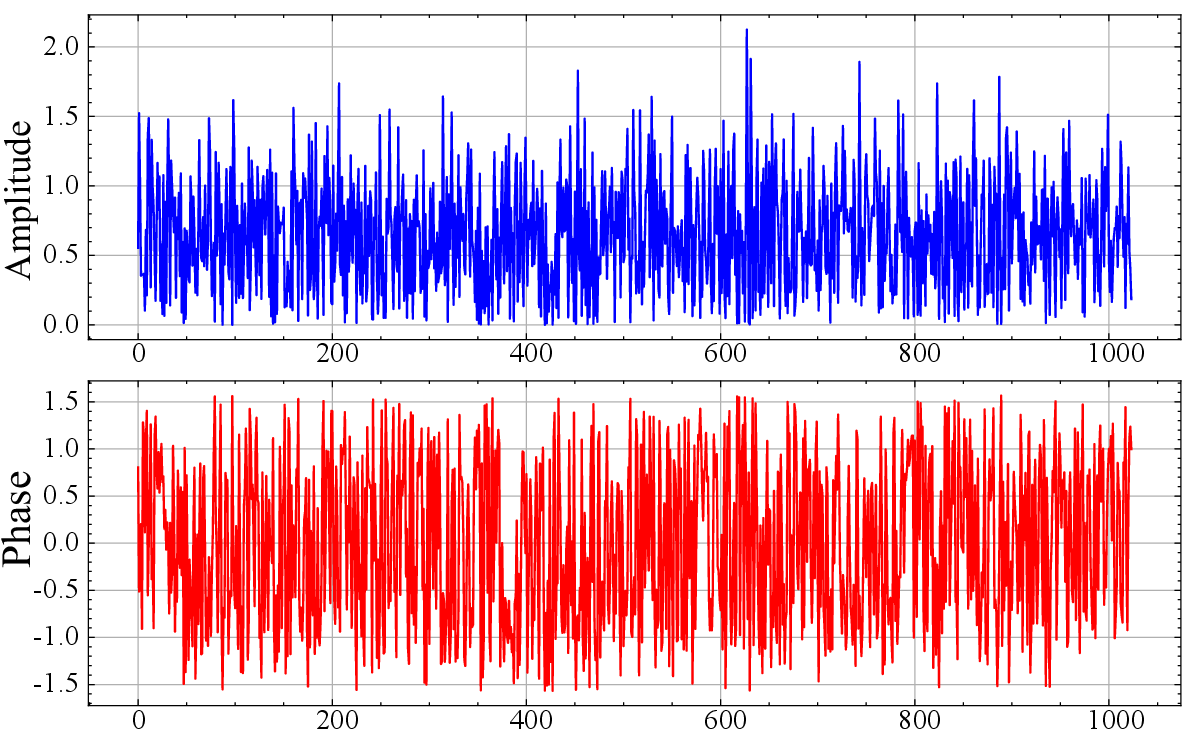}}
\subfigure[]{\includegraphics[height=3.5cm]{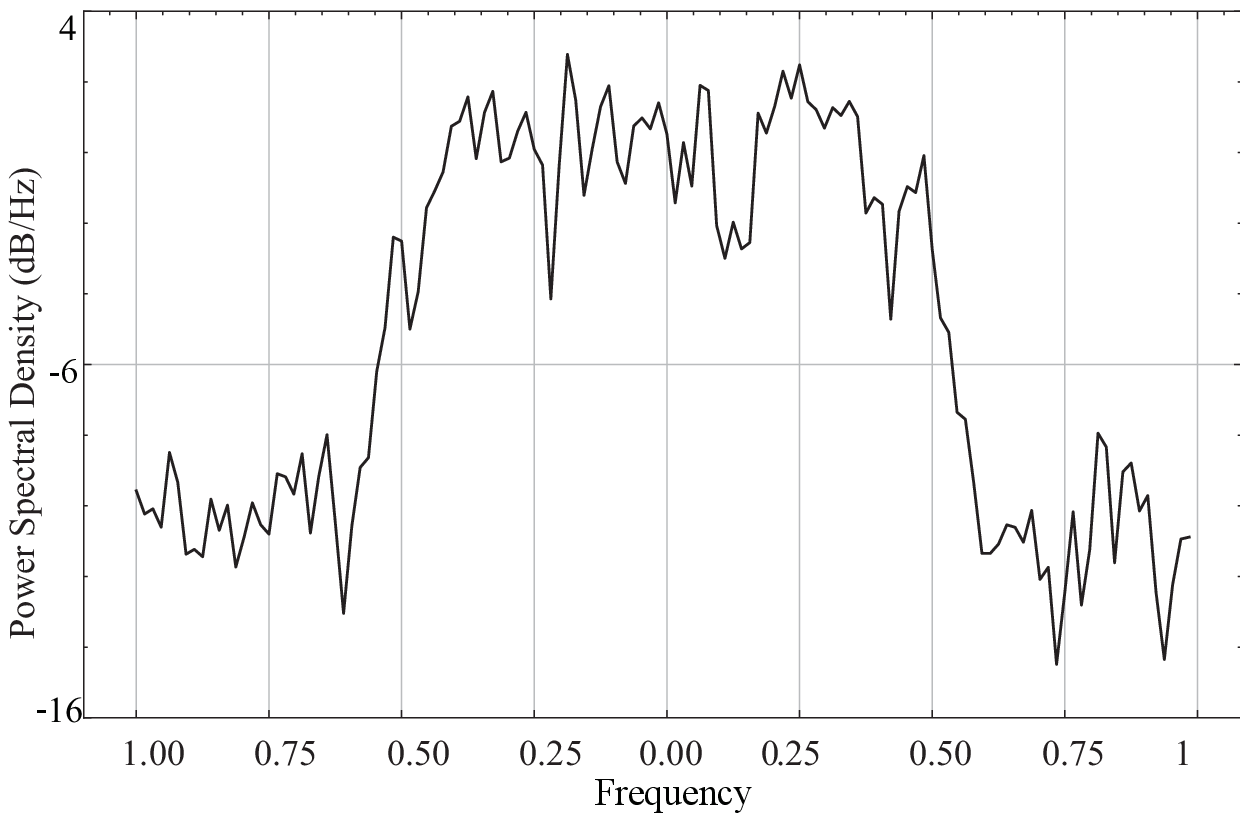}}
\DeclareGraphicsExtensions.
\caption{Different representations of the received signal of HisarMod2019.1 \cite{tekbiyik2020robust}, including (a) IQ signal, (b) amplitude and phase (AP), and (c) power spectral density (PSD).}
\label{fig:data_representation2019}
\end{figure*}

The input signal can be represented in different ways, such as IQ signal representation, amplitude and phase (AP) representation, and power spectral density (PSD) representation, as shown in Fig. \ref{fig:data_representation2016} and Fig. \ref{fig:data_representation2019}.

\paragraph{IQ signal representation} The IQ signal can be represented as in Eq. \ref{eq:iq_signal}.

\paragraph{Amplitude and phase (AP) representation}

The IQ signal representation can be converted into an amplitude ($A$) and phase ($P$) representation using the following formulas. 
Amplitude (A): This is calculated as the magnitude of the vector represented by the I and Q components. Mathematically, it is given by
\begin{equation}
	A = \sqrt{I^2 + Q^2}.
\end{equation}

Phase ($P$)  is the angle of the vector concerning the in-phase axis. It can be calculated using the arctangent function, specifically using the two-argument variant to take into account the correct quadrant of the angle. The formula is
\begin{equation}
	P = \arctan\left(\frac{Q}{I}\right).
\end{equation}

\paragraph{Power spectral density (PSD) representation}
PSD representation is a fundamental tool in signal processing and communications, providing a detailed view of a signal's power distribution across its frequency components. The transformation of IQ to PSD provides valuable insights into the frequency-domain characteristics of complex baseband signals. The PSD, $S_{xx}(f)$, of a wide-sense stationary random process $x(t)$ is formally defined as the Fourier transform of its autocorrelation function $R_{xx}(\tau)$
\begin{equation}
S_{xx}(f) = \int_{-\infty}^{\infty} R_{xx}(\tau) e^{-j2\pi f\tau} d\tau.
\end{equation}

In practice, for discrete-time IQ signals, the PSD is often estimated using Welch's method, which offers a balance between resolution and statistical variance. This method segments the signal, computes modified periodograms of these segments, and averages them to produce the PSD estimate
\begin{equation}
\hat{P}{xx}(f) = \frac{1}{K} \sum{k=1}^{K} |\hat{X}_k(f)|^2,
\end{equation}
where $K$ is the number of segments and $\hat{X}_k(f)$ is the discrete Fourier transform of the $k$-th windowed segment.
	
\subsection{Multi-Sequence Multi-Scale Attention Feature Extractor}
The multi-sequence multi-scale attention feature extractor (MS-MSANet) is used to extract the features from the input signal.
Multi-scale module has been demonstrated useful for signal recognition tasks \cite{zhang2021novel,yuan2021multiscale,ding2022data}. 
In this work, we introduce a multi-scale attention network (MSANet) to further enhance the feature representation. 
The MSANet consists of two layers of multi-scale attention modules, which is composed of a multi-scale module and a channel attention module. 
The multi-scale module is designed to capture the multi-scale features of the input signal, learning more separable features. 
Moreover, the channel attention mechanism is introduced to model the channel-wise dependencies and enhance the important features. 
Finally, a global average pooling layer is used to aggregate the feature maps and an $L_2$ normalization layer is used to normalize the feature vectors. 

In the multi-scale module, a convolutional layer with kernel $3\times 1$, stride $2$ is used to downsample the input signal, labeling it as $C_{\text{down}}$. 
Then, the downsampled signal is fed into four parallel convolutional layers with different kernel sizes. 
The four parallel convolutional layers consist of $1 \times 1$, $3 \times 1$, $5 \times 1$, and $7 \times 1$ kernels, respectively, and the convolutional operation is represented as $Conv_1, Conv_3, Conv_5, Conv_7$. 
At the beginning of the $Conv_3, Conv_5, Conv_7$ layers, a $1 \times 1$ convolutional layer is used to reduce the channel dimension, denoting $C_1$. 
All the convolutional layers are followed by a batch normalization layer and a ReLU activation function. 
The output feature maps of the four convolutional layers are concatenated along the channel dimension. 
The multi-scale module can be formulated as 
\vspace{-2pt}
\begin{equation}
	\boldsymbol{X}_{\text{multi}} = 
	\text{Concat}(\boldsymbol{X}_{1}, \boldsymbol{X}_{2}, \boldsymbol{X}_{3}, \boldsymbol{X}_{4}),
\end{equation}
where $\boldsymbol{X}_{1}$, $\boldsymbol{X}_{2}$, $\boldsymbol{X}_{3}$, and $\boldsymbol{X}_{4}$ are the output feature maps of the four convolutional layers, given as
 \begin{subequations}
 \begin{align}
	\boldsymbol{X}_{1} &= Conv_1(C_{\text{down}}(\boldsymbol{X})),\\
	\boldsymbol{X}_{2} &= Conv_3(C_1(C_{\text{down}}(\boldsymbol{X}))),\\
	\boldsymbol{X}_{3} &= Conv_5(C_1(C_{\text{down}}(\boldsymbol{X}))),\\
	\boldsymbol{X}_{4} &= Conv_7(C_1(C_{\text{down}}(\boldsymbol{X}))).
 \end{align}
\end{subequations}

\begin{figure}
	\centering
	\includegraphics[width=0.95\linewidth]{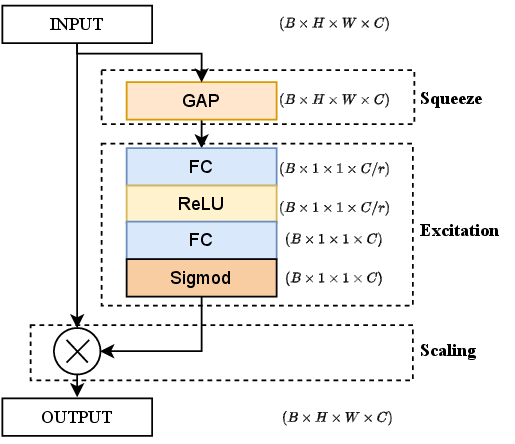} 
	\caption{Squeeze-and-Excitation Module.}
	\label{fig:se}
\end{figure}

To further enhance the feature representation, we introduce the channel attention mechanism into the feature extraction module. 
The channel attention mechanism can capture the inter-channel dependencies and enhance the important features. The squeeze-and-excitation (SE) module \cite{hu2018squeeze} is adopted to model the channel-wise dependencies. 
As shown in Fig. \ref{fig:se}, the SE module consists of two operations, namely, squeeze and excitation. 
The goal of the squeeze operation is to aggregate spatial information within each channel to produce a channel descriptor. This is typically achieved through global average pooling, which reduces the spatial dimensions (height and width) of each channel, retaining only the channel dimension. Mathematically, for an input feature map $\boldsymbol{X} \in \mathbb{R}^{H\times W\times C}$, the squeeze operation can be expressed as
\begin{equation}
	\boldsymbol{z}_c = \text{GAP}(\boldsymbol{X}) = \frac{1}{H\times W} \sum_{i=1}^{H} \sum_{j=1}^{W} \boldsymbol{x}^c_{i,j},
\end{equation}
where $z_c$ is the feature obtained by performing global average pooling over the $c$-th channel, and $\boldsymbol{x}^c_{i,j}$ is the value of the $c$-th channel at position $(i, j)$. 

The excitation operation utilizes the channel descriptors generated in the squeeze operation to learn a channel-specific weight vector. This is typically implemented using a simple fully connected layer, including an ReLU activation followed by a Sigmoid activation to generate the weights. Mathematically, the excitation function can be represented as
\begin{equation}
	\boldsymbol{s} = \sigma(\boldsymbol{W}_2 \delta(\boldsymbol{W}_1 \boldsymbol{z})),
\end{equation}
where $\sigma$ and $\delta$ are the Sigmoid and ReLU activation functions, respectively. $\boldsymbol{W}_1 \in \mathbb{R}^{\frac{C}{r}\times C}$ and $\boldsymbol{W}_2 \in \mathbb{R}^{C\times \frac{C}{r}}$ are the weights of the fully connected layers, and $r$ is the reduction ratio. 

Finally, the obtained weights $\boldsymbol{s}$ are used to scale the original input feature map $\boldsymbol{X}$ across channels by performing channel-wise multiplication
\begin{equation}
	\boldsymbol{X}_{\text{SE}} = \boldsymbol{X} \cdot \boldsymbol{s},
\end{equation}
where $\boldsymbol{X}_{\text{SE}}$ is the output feature map after the SE module.

\subsection{Meta-Protype Training}
As shown in Fig. \ref{fig:framework}, the feature extractor is trained using a supervised, episode-based methodology. 
The feature encoder MSANet, denoted as $f(\cdot)$, is systematically trained utilizing a supervised, episode-based methodology. 
Each episode commences with the data loader introducing a batch of training data procured from the main dataset. This batch is distinctively composed, ensuring an equitable representation of samples from $N$ distinct classes. 

Considering an episodic batch that includes $S$ support samples $\{x_{i,j}^S\}_{i=1}^S$ and $Q$ query samples $\{x_{i,j}^Q\}_{i=1}^Q$ from $N$ classes, for a total of $(S+Q) \times N$ support and query samples per batch. At every episode, a set of prototypes $c=\{c_j\}_{j=1}^N$ is firstly computed, given as
\begin{equation}
	c_j = \frac{1}{S} \sum_{i=1}^S f(x_{i,j}^S).
\label{eq:prototype}
\end{equation}

During the training process, a triplet loss is used to minimize the distance between the query sample and the correct class prototype, while maximizing the distance between the query sample and the incorrect class prototypes. 
The triplet loss can be formulated as
\begin{equation}
	\mathcal{L} =\frac{1}{N_t} \sum_{i=1}^{N_t} \max(0, \textbf{d}(x_{i,j}^{Q^{+}},c_j) - \textbf{d}(x_{i,j}^{Q^{-}},c_k) + m),	
\end{equation}
where $x_{i,j}^{Q^{+}}$ and $x_{i,j}^{Q^{-}}$ are the query samples from the correct and incorrect classes, respectively. $m$ is the margin parameter, and $N_t$ is the number of triplets. $\textbf{d}$ is the distance function, given as
\begin{equation}
	\textbf{d}(x,j) = -d_{L2}(f(x), c_j),
\label{eq:distance}
\end{equation}
and $-d_{L2}$ is the Euclidean distance function.

\subsection{Modular Open-Set Classifier}
At inference time, the FSOS-AMC pipeline receives $K$ enrollment samples for every user-defined modulation scheme. After computing the embeddings of the $K$-shot samples using the trained encoder, a classifier computes Eq. \ref{eq:prototype} and stores the class prototypes. 

When a new test sample is fed to the pipeline, the open-set classifier returns a probability score vector $P = \{p_i\}_{i=0}^N$ based on the current prototype set. 
Specifically, $p_0$ is the prediction score for the unknown class and $p_{i}$ is the probability of the $i$-th class, which assumes the highest value if the distance score of Eq. \ref{eq:distance} is the lowest. The final class prediction $y$ is
\begin{equation}
y=\begin{cases}
		\arg \max p_i,\ \mathrm{if }\ \ p_i \ge \gamma\\
	   0,\qquad \qquad  \mathrm{otherwise}
	   \end{cases}
\end{equation}
where $y = 0$ denotes the unknown class and $\gamma$ is a manually tunable parameter to tradeoff between the classifier’s precision and recall. 

In this work, the Nearest Class Mean (NCM) classifier \cite{hayes2022online} is adopted as the open set classifier, which computes the distance between the query sample and the class prototypes. 
A simple open-set variant estimates the $c_0$ prototype for the unknown class using $K$ random samples taken from the target domain but not belonging to the pre-defined classes. 
The probability score is computed by applying the SoftMax on the ($N+1$)-sized distance vector obtained from Eq. \ref{eq:distance}.

\section{Experiments and Results}
\label{sec:experiments_results}

\subsection{Evaluation Metrics}
Open set recognition is a challenging problem in DL-based recognition tasks. 
To better understand the difficulty of the open set recognition problem, the \emph{openness}  \cite{scheirer2012toward} metric is defined as 
\begin{equation}
	openness = 1 - \sqrt{\frac{2\times N_{tr}}{N_{tr}+N_{te}}},
\end{equation}
where $N_{tr}$ represents the number of known modulation schemes in the training set, and $N_{te}$ is the number of known modulation schemes in the testing set. Typically, the openness value is between 0 and 1. A higher openness value indicates a more challenging open set recognition problem. 
$N_{tr} \leq N_{te}$. When $N_{tr} = N_{te}$, the value of $openness$ is 0 and it is closed set recognition. 

To evaluate the performance of the proposed FSOS-AMC framework, the classification accuracy is used to evaluate the performance of the proposed FSOS-AMC framework for known classes. 
Moreoverm, to evaluate the performance of the proposed FSOS-AMC framework for unknown classes, the metric \emph{AUROC} (Area Under the Receiver Operating Characteristic curve) is used. 
The larger the \emph{AUROC} value, the better the performance of the open set recognition. 

\subsection{Experimental Setup}

To evaluate the performance of our proposed FSOS-AMC framework, we conduct experiments on the RadioML 2016.10a dataset \cite{o2016convolutional}, which is a widely used dataset for AMC. The dataset contains 11 modulation schemes, including 8 digital modulation schemes and 3 analog modulation schemes at varying signal-to-noise ratios from $-20$dB to $18$dB with a step of $2$. 
The digital modulation schemes include 8PSK, BPSK, CPFSK, GFSK, PAM4, QAM16, QAM64, QPSK. The analog modulation schemes include AM-DSB, AM-SSB, and WBFM. 
The dataset is generated by GNU Radio and is available on the website\footnote{\url{https://www.deepsig.ai/datasets}}. 
By the specifications outlined in \cite{zhang2024sswsrnet}, we opt for a signal-to-noise ratio (SNR) range of -6dB to 12 dB with an interval of 2 dB in our experiment, with total $110,000$ samples.
The proportion of the training set and testing set is $80\%$ and $20\%$, respectively. 
To perform open set recognition, we randomly select 6 modulation schemes (AM-DSB, QAM64, CPFSK, GFSK, 8PSK, PAM4) as the training set and all 11 modulation schemes as the testing set, where 5 modulation schemes are unknown during testing. 
For this setting, the \emph{openness} value is $15.98\%$. 
The dataset configuration is shown in Table \ref{tab:dataset}, the compared methods and the proposed FSOS-AMC framework are first trained on the 5 known modulation schemes using a 5-way 10-shot learning manner, and then tested on the 11 modulation schemes. 

To investigate the performance of the proposed FSOS-AMC framework under multipath fading channels, we conduct experiments on the HisarMod2019.1 dataset \cite{tekbiyik2020robust}. 
The dataset is a diverse signal dataset created to enhance research in wireless communication and modulation classification. It comprises $780,000$ signals across 26 modulation classes from 5 modulation families (analog, FSK, PAM, PSK, and QAM), as shown in Table \ref{tab:modulation_types}. Each signal consists of 1024 I/Q samples, generated using MATLAB 2017a with an oversampling rate of 2 and raised cosine pulse shaping. The dataset covers 20 SNR levels from -20dB to 18dB and simulates 5 different wireless communication channels (ideal, static, Rayleigh, Rician, and Nakagami-m), with equal distribution among these channels. This comprehensive dataset aims to provide a robust resource for researchers, comparable to the RadioML2016.10a dataset, and includes various fading models to represent real-world communication scenarios, as shown in Table \ref{table_comparison}.
The dataset configuration is shown in Table \ref{tab:dataset2019}, the compared methods and the proposed FSOS-AMC framework are first trained on the 13 known modulation schemes using a 5-way 10-shot learning manner, and then tested on the 26 modulation schemes.

\begin{table}[!h]
	\caption{26 Different Modulation Types in Hisarmod2019.1}
	\centering
	\begin{tabular}{m{3cm}<{\centering}m{4.5cm}<{\centering}}
		\toprule
		Modulation Family & Modulation Types \\
		\midrule
		Analog & AM-DSB, AM-SC, AM-USB, AM-LSB, FM, PM \\
		\midrule
		FSK & 2-FSK, 4-FSK, 8-FSK, 16-FSK \\
		\midrule
		PAM & 4-PAM, 8-PAM, 16-PAM \\
		\midrule
		PSK & BPSK, QPSK, 8-PSK, 16-PSK, 32-PSK, 64-PSK \\
		\midrule
		QAM & 4-QAM, 8-QAM, 16-QAM, 32-QAM, 64-QAM, 128-QAM, 256-QAM \\
		\bottomrule
\end{tabular}
\label{tab:modulation_types}
\end{table}

\begin{table}[!t]
\renewcommand{\arraystretch}{1.3}
\caption{Comparison of HisarMod2019.1 \cite{tekbiyik2020robust} and RadioML2016.10a}
\label{table_comparison}
\centering
\begin{tabular}{m{=1.8cm}<{\centering}m{2.7cm}<{\centering}m{2.7cm}<{\centering}}
\toprule
\textbf{Feature} & \textbf{HisarMod2019.1} & \textbf{RadioML2016.10a} \\
\midrule
Total Signals & 780,000 & 220,000 \\
\midrule
Mod. Classes & 26 & 11 \\
\midrule
Mod. Families & Analog, FSK, PAM, PSK, QAM &8 digital, 3 analog\\
\midrule
Signals/Mod. & 1,500 & 1,000 \\
\midrule
Signal Length & 1,024 I/Q samples & 128 I/Q samples \\
\midrule
SNR Levels & 20 (-20dB to 18dB) & 20 (-20dB to 18dB) \\
\midrule
Channels & Ideal, Static, Rayleigh, Rician (k=3), Nakagami-m (m=2) & Fading channel with AWGN noise\\
\midrule
Signals/Channel & 300 & - \\
\midrule
Oversamp. Rate & 2 & - \\
\midrule
Pulse Shaping & Raised cosine (0.35) & - \\
\midrule
Multipath Taps & 4 or 6 & - \\
\midrule
Gen. Tool & MATLAB 2017a & GNU Radio\\
\bottomrule
\end{tabular}
\end{table}

\begin{table}[]
\centering
\caption{Dataset Configuration for Few-Shot Open-Set AMC Using RadioML 2016.10a Dataset.}
\begin{center}
\begin{tabular}{m{=0.5cm}<{\centering}m{3.5cm}<{\centering}m{3.5cm}<{\centering}}
\toprule
& Training & Testing\\
\midrule
Classes & AM-DSB, QAM64, CPFSK,GFSK, 8PSK, PAM4 
& AM-DSB, QAM64, CPFSK,GFSK, 8PSK, PAM4, unknown: \textbf{\{\emph{AM-SSB, BPSK, QAM16, QPSK, WBFM}\}} 
 \\
\midrule
Numbers & 5-way 10-shot& 
$14,000$ samples\\
\midrule
\emph{openness} & \multicolumn{2}{c}{$15.98\%$}  \\
\bottomrule
\end{tabular}
\label{tab:dataset}
\end{center}
\end{table}	

\begin{table}[]
	\centering
	\caption{Dataset Configuration for Few-Shot Open-Set AMC Using HisarMod2019.1 Dataset.}
	\begin{center}
	\begin{tabular}{m{=0.5cm}<{\centering}m{3.5cm}<{\centering}m{3.5cm}<{\centering}}
	\toprule
	& Training & Testing\\
	\midrule
	Classes & 64PSK, AM-DSB, FM, 16PSK, BPSK, 8FSK, 2FSK, 4FSK, QPSK, 128QAM, 4PAM, 8PAM, 64QAM
	& 64PSK, AM-DSB, FM, 16PSK, BPSK, 8FSK, 2FSK, 4FSK, QPSK, 128QAM, 4PAM, 8PAM, 64QAM, unknown: \textbf{\{\emph{8PSK, 32PSK, 4QAM, 8QAM, 16QAM, 32QAM, 256QAM, 16FSK, 16PAM, AM-DSB-SC, AM-USB, AM-LSB, PM}\}} 
	 \\
	\midrule
	Numbers & 5-way 10-shot& 
	$130,000$ samples\\
\midrule
\emph{openness} & \multicolumn{2}{c}{$18.35\%$}  \\
\bottomrule
\end{tabular}
\label{tab:dataset2019}
\end{center}
\end{table}	

Experiments are conducted on a workstation with an AMD Ryzen 9 7900X3D CPU, 64GB RAM, and an NVIDIA GeForce RTX 4070 Ti GPU (12GB VRAM), and the system is running Ubuntu 22.04. 
Code is implemented in Python and the PyTorch deep learning framework \cite{paszke2019pytorch}. 
To train our proposed FSOS-AMC framework, we employ the Adam optimization algorithm with an initial learning rate of $0.001$ over $50$ training epochs. 
During each training epoch, $200$ episodes of $5$ modulation schemes with $10$ samples from each modulation scheme are selected. 

\subsection{Performance Comparison with State-of-the-Art Methods under Multipath Fading Channels}
To demonstrate the effectiveness of the proposed FSOS-AMC framework, we first compare it with the state-of-the-art methods in terms of accuracy, including CNN \cite{o2016convolutional}, ResNet \cite{o2018over}, MSNet \cite{zhang2021novel}, and SSwsrNet \cite{zhang2024sswsrnet} using the RadioML 2016.10A dataset \cite{o2016convolutional}. 
The proposed MS-MSANet model consistently outperforms the other methods across all SNR levels, achieving the highest accuracy of 82.81\%. It is followed closely by MSANet-IQ (80.99\%) and SSwsrNet (80.55\%). MSNet, ResNet, and VGG show lower performance, with VGG having the lowest overall accuracy at 70.97\%. All models demonstrate improved accuracy as SNR increases, with the most significant gains observed between -6 dB and 0 dB. The performance curves tend to plateau or show minimal improvements beyond 6 dB SNR for most models. This comparison highlights the robustness and superior performance of the MS-MSANet model in classification tasks under varying noise conditions within the FSOS-AMC framework.

\begin{figure}
\centering
\includegraphics[width=0.95\linewidth]{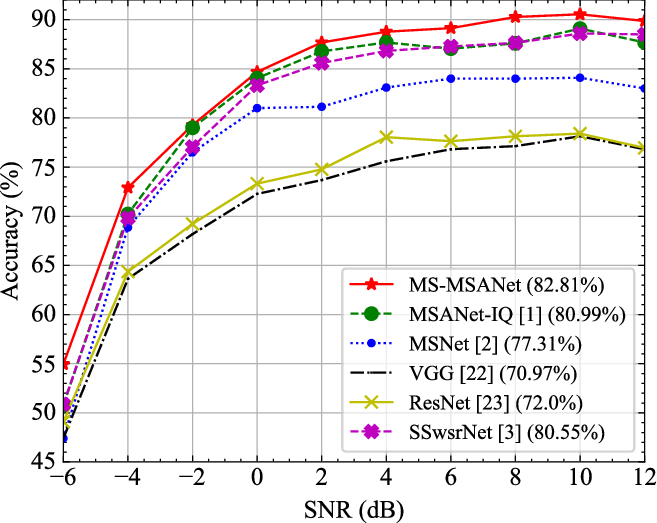} 
\caption{Classification performance comparison of the proposed FSOS-AMC framework with the state-of-the-art methods, including VGG \cite{o2016convolutional}, ResNet \cite{o2018over}, MSNet \cite{zhang2021novel}, and SSwsrNet \cite{zhang2024sswsrnet} under various signal-to-noise ratios (SNR) using RadioML 2016.10A \cite{o2016convolutional}.}
\label{fig:comparison}
\end{figure}

\begin{figure}
\centering
\subfigure[]{
\includegraphics[width=0.95\linewidth]{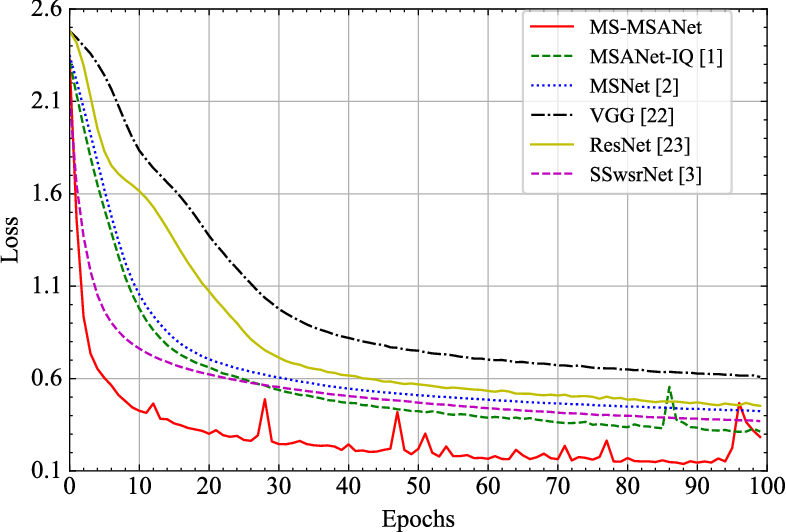}}
\subfigure[]{
\includegraphics[width=0.95\linewidth]{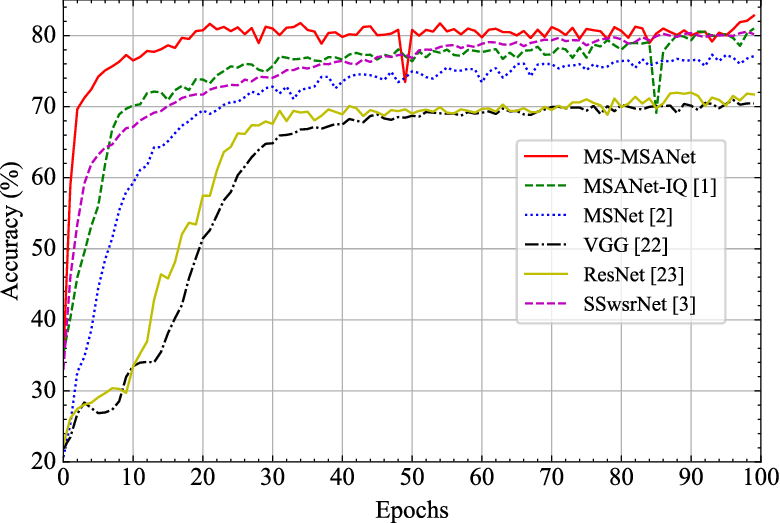}}
\DeclareGraphicsExtensions.
\caption{The comparison of training process of the propsoed FSOS-AMC framework with the state-of-the-art methods, including CNN \cite{o2016convolutional}, ResNet \cite{o2018over}, MSNet \cite{zhang2021novel}, and SSwsrNet \cite{zhang2024sswsrnet} in terms of (a) training loss and (b) test accuracy using RadioML 2016.10A \cite{o2016convolutional}.}
\label{fig:comparison_training}
\end{figure}

To better understand the training process of the proposed FSOS-AMC framework, we compare the training process including the training loss and the test accuracy of the proposed FSOS-AMC framework ((MS-MSANet)) with the state-of-the-art methods, including CNN \cite{o2016convolutional}, ResNet \cite{o2018over}, MSNet \cite{zhang2021novel}, and SSwsrNet \cite{zhang2024sswsrnet} in terms of training loss and test accuracy using RadioML 2016.10A \cite{o2016convolutional}, as shown in Fig. \ref{fig:comparison_training}. 
As can be observed, the proposed FSOS-AMC framework exhibits superior performance with the fastest convergence, lowest loss, and highest accuracy (82.81\%). Notably, MSANet-IQ demonstrates strong performance as the second-best model, showing rapid initial improvement in both loss reduction and accuracy gain. It maintains consistent second-place performance throughout training, achieving approximately 80.99\% accuracy by the end, and outperforms VGG, ResNet, MSNet, and SSwsrNet across all epochs. 
With the multi-sequence signal representation, the proposed FSOS-AMC framework can reach the highest test accuracy with about $20$ epochs, while the other models need more epochs to achieve the highest test accuracy. 

\begin{figure}
\centering
\includegraphics[width=0.95\linewidth]{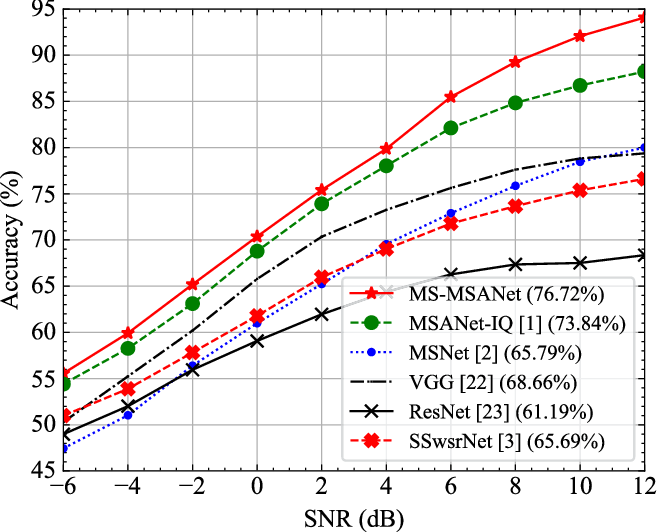} 
\caption{Classification performance comparison of the proposed FSOS-AMC framework with the state-of-the-art methods, including VGG \cite{o2016convolutional}, ResNet \cite{o2018over}, MSNet \cite{zhang2021novel}, and SSwsrNet \cite{zhang2024sswsrnet} under various signal-to-noise ratios (SNRs) using HisarMod2019.1 \cite{tekbiyik2020robust} under multipath fading channels.}
\label{fig:h2019_comparison}
\end{figure}

To further validate the effectiveness of the proposed FSOS-AMC framework especially under multipath fading channels, we compare the performance of the proposed FSOS-AMC framework with the state-of-the-art methods, including CNN \cite{o2016convolutional}, ResNet \cite{o2018over}, MSNet \cite{zhang2021novel}, and SSwsrNet \cite{zhang2024sswsrnet} under various signal-to-noise ratio (SNR) levels using HisarMod2019.1 \cite{tekbiyik2020robust}, as shown in Fig. \ref{fig:h2019_comparison}. 
The proposed FSOS-AMC framework, represented by MS-MSANet and MSANet-IQ, demonstrates superior performance compared to VGG, ResNet, MSNet, and SSwsrNet. MS-MSANet consistently outperforms all other models across all SNR levels, achieving the highest accuracy of 76.72\%, followed by MSANet-IQ at 73.84\%. All models show improved performance as SNR increases, indicating better classification in less noisy conditions. The experiment, conducted using HisarMod2019.1 under multipath fading channels, provides a realistic scenario for evaluating these models in challenging signal environments. ResNet shows the lowest overall performance among the compared models, while the FSOS-AMC framework demonstrates its effectiveness in handling various noise levels in signal classification tasks. 

\begin{figure}
	\centering
	\subfigure[]{
	\includegraphics[width=0.95\linewidth]{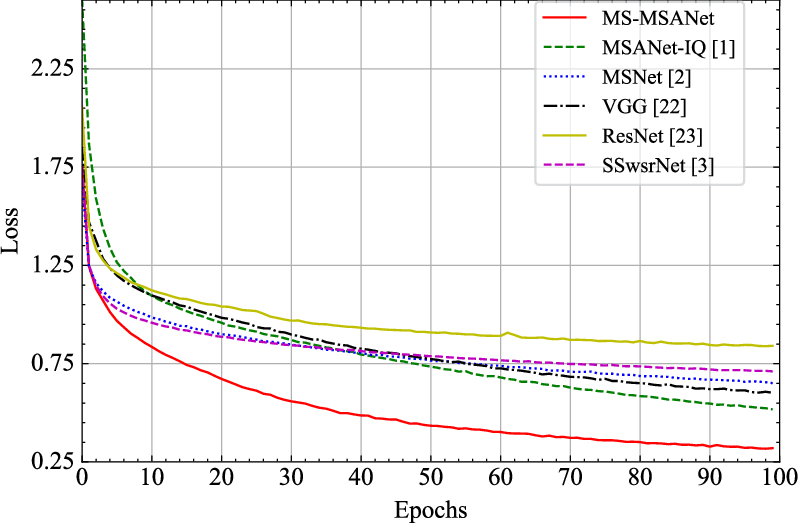}}
	\subfigure[]{
	\includegraphics[width=0.95\linewidth]{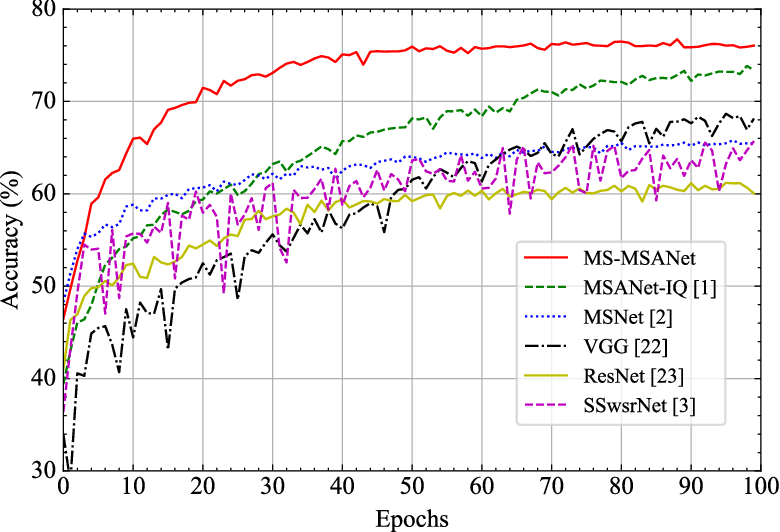}}
	\DeclareGraphicsExtensions.
	\caption{The comparison of training process of the proposed FSOS-AMC framework with the state-of-the-art methods, including CNN \cite{o2016convolutional}, ResNet \cite{o2018over}, MSNet \cite{zhang2021novel}, and SSwsrNet \cite{zhang2024sswsrnet} in terms of (a) training loss and (b) test accuracy using HisarMod2019.1 \cite{tekbiyik2020robust} under multipath fading channels.}
\label{fig:h19_comparison_training}
\end{figure}

Fig. \ref{fig:h19_comparison_training} illustrates the training process of the proposed FSOS-AMC framework with the state-of-the-art methods, including VGG \cite{o2016convolutional}, ResNet \cite{o2018over}, MSNet \cite{zhang2021novel}, and SSwsrNet \cite{zhang2024sswsrnet} using HisarMod2019.1 \cite{tekbiyik2020robust} under multipath fading channels. 
In Fig. \ref{fig:h19_comparison_training} (a), MS-MSANet demonstrates the lowest and most rapidly decreasing loss curve, indicating superior learning efficiency. Fig. \ref{fig:h19_comparison_training} (b) reveals that MS-MSANet achieves and maintains the highest accuracy throughout the training process, reaching approximately 75-80\% by the 100th epoch. MSANet-IQ performs second-best in both loss reduction and accuracy gain. The other models (MSNet, VGG, ResNet, and SSwsrNet) show varying degrees of performance but consistently lag behind the top two. This comprehensive comparison highlights the effectiveness of the FSOS-AMC framework, particularly the MS-MSANet variant, in achieving lower training loss and higher test accuracy compared to existing methods in the challenging scenario of multipath fading channels.

\subsection{Performance using Different Signal Representations}

\begin{figure}
\centering
\includegraphics[width=0.95\linewidth]{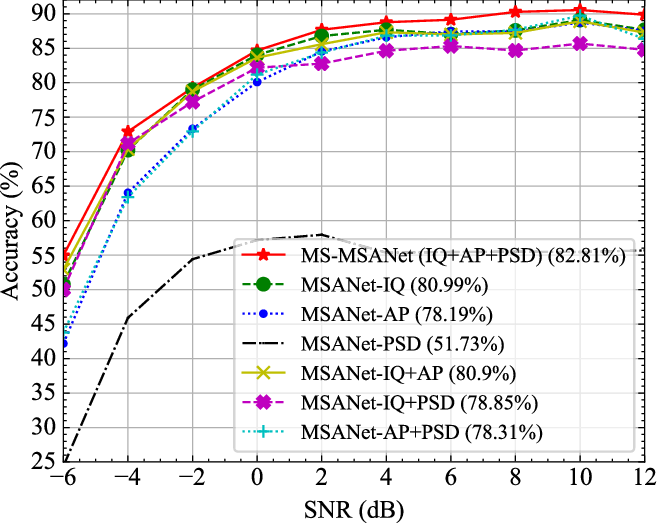} 
\caption{Classification performance comparison of the proposed FSOS-AMC framework using different signal representations, including IQ signal, amplitude and phase, and power spectral density (PSD) under various signal-to-noise ratios (SNRs) using RadioML 2016.10A \cite{o2016convolutional}.}
\label{fig:data_representation}
\end{figure}

To demonstrate the impact of different signal representations on the classification performance of the proposed FSOS-AMC framework, we compare the performance of the proposed FSOS-AMC framework using different signal representations, including IQ signal, amplitude and phase (AP), and power spectral density (PSD) under various signal-to-noise ratios (SNRs) using RadioML 2016.10A \cite{o2016convolutional}. 
Fig. \ref{fig:data_representation} illustrates the performance of various MSANet configurations across different SNR levels. The MS-MSANet (IQ+AP+PSD) model demonstrates superior performance, achieving the highest overall accuracy of 82.81\%. This is followed closely by MSANet-IQ and MSANet-IQ+AP, both attaining accuracies around 81\%. The MSANet-AP, MSANet-IQ+PSD, and MSANet-AP+PSD models show comparable performance, with accuracies ranging from 78.19\% to 78.85\%. Notably, the MSANet-PSD model exhibits significantly lower accuracy at 51.73\%. All models display a general trend of improved accuracy as SNR increases, with the most substantial gains observed between -6 dB and 0 dB. The performance curves tend to plateau at higher SNR levels, particularly above 6 dB.

\begin{figure}
\centering
\subfigure[]{
\includegraphics[width=0.95\linewidth]{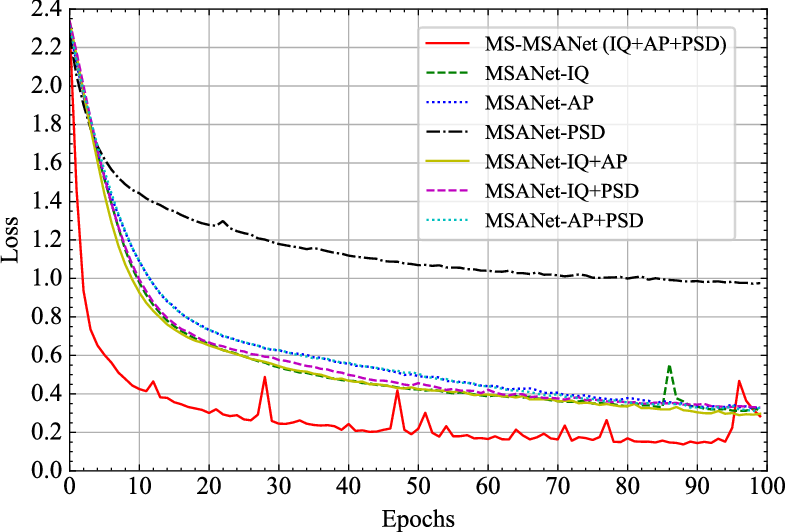}}
\subfigure[]{
\includegraphics[width=0.95\linewidth]{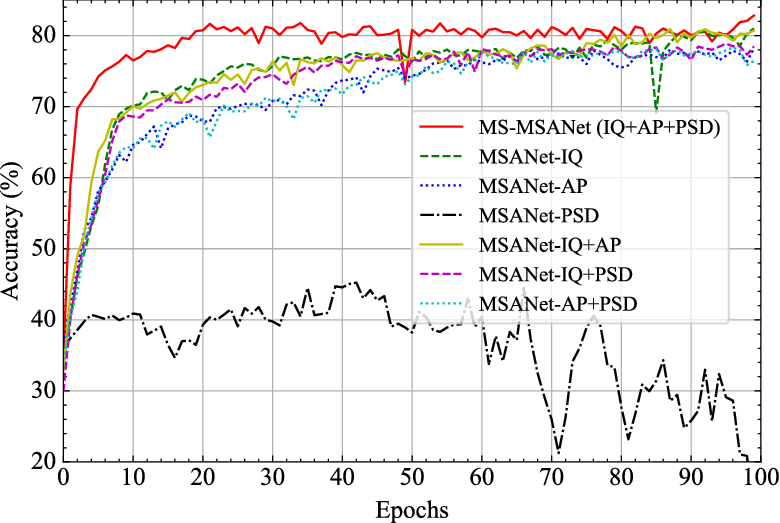}}
\DeclareGraphicsExtensions.
\caption{The comparison of training process of the proposed FSOS-AMC framework using different signal representations, including IQ signal, amplitude and phase, and power spectral density (PSD) in terms of (a) training loss and (b) test accuracy using RadioML 2016.10A \cite{o2016convolutional}.}
\label{fig:data_training}
\end{figure}

Fig. \ref{fig:data_training} shows the training process of the proposed FSOS-AMC framework using different signal representations, including IQ signal, amplitude and phase, and power spectral density (PSD) in terms of training loss and test accuracy using RadioML 2016.10A \cite{o2016convolutional}, where Fig. \ref{fig:data_training} (a) shows the training loss and Fig. \ref{fig:data_training} (b) shows the test accuracy. 
It can be observed that the proposed FSOS-AMC framework using multi-sequence signal representation achieves the lowest training loss and the highest test accuracy, followed by the amplitude and phase representation. The PSD representation shows the highest training loss and the lowest test accuracy.

\begin{figure}
	\centering
	\includegraphics[width=0.95\linewidth]{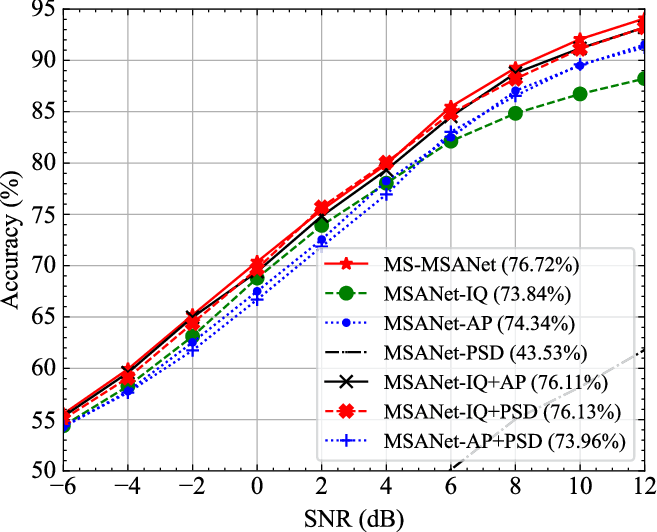} 
	\caption{Classification performance comparison of the proposed FSOS-AMC framework using different signal representations, including IQ signal, amplitude and phase, and power spectral density (PSD) under various signal-to-noise ratio (SNR) using HisarMod2019.1 \cite{tekbiyik2020robust} under multipath fading channels.}
	\label{fig:h19_representation}
\end{figure}

To further validate the effectiveness of the proposed FSOS-AMC framework using different signal representations, we compare the performance of the proposed FSOS-AMC framework using different signal representations, including IQ signal, amplitude and phase, and power spectral density (PSD) under various signal-to-noise ratios (SNRs) using HisarMod2019.1 \cite{tekbiyik2020robust} under multipath fading channels, as shown in Fig. \ref{fig:h19_representation}. 
It can be seen that all techniques generally improve in accuracy as SNR increases, with MS-MSANet (with IQ+AP+PSD) and MSANet-IQ+PSD performing best overall. MSANet-PSD notably underperforms compared to the others. The performances tend to converge at higher SNR values, particularly above 8 dB. This visualization is crucial for researchers in signal processing and modulation classification, offering insights into the effectiveness of different signal representation techniques under varying noise conditions.

\subsection{Performance Comparison under Few-Shot Open-Set Scenarios} 
\begin{figure}
\centering
\includegraphics[width=0.95\linewidth]{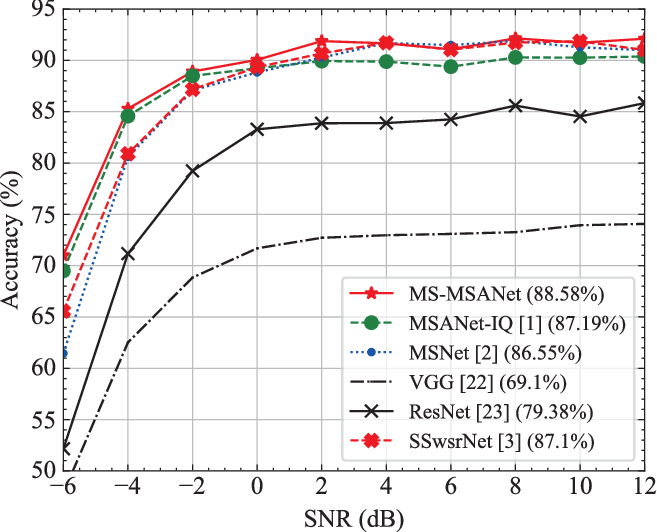} 
\caption{Classification performance comparison of the proposed FSOS-AMC framework under various signal-to-noise ratios (SNRs) using RadioML 2016.10A \cite{o2016convolutional} under \emph{few-shot open-set} scenarios.}
\label{fig:fsos_2016}
\end{figure}

To evaluate the performance of the proposed FSOS-AMC framework under \emph{few-shot open-set} scenarios, we compare the classification performance of the proposed FSOS-AMC framework with the state-of-the-art methods, including CNN \cite{o2016convolutional}, ResNet \cite{o2018over}, MSNet \cite{zhang2021novel}, and SSwsrNet \cite{zhang2024sswsrnet} under various signal-to-noise ratio (SNR) levels using RadioML 2016.10A \cite{o2016convolutional}, as shown in Fig. \ref{fig:fsos_2016}. 
It can be found that MS-MSANet consistently outperforms other models across most SNR levels, achieving the highest accuracy of 88.58\%. MSANet-IQ and SSwsrNet show comparable performance, with accuracies around 87\%. MSNet and ResNet demonstrate lower performance, with VGG having the lowest overall accuracy at 78.21\%. The proposed FSOS-AMC framework shows superior performance in classifying known and unknown modulation schemes under \emph{few-shot open-set} scenarios, highlighting its robustness and effectiveness in handling challenging classification tasks.

\begin{figure}[!h]
\centering
\includegraphics[width=0.95\linewidth]{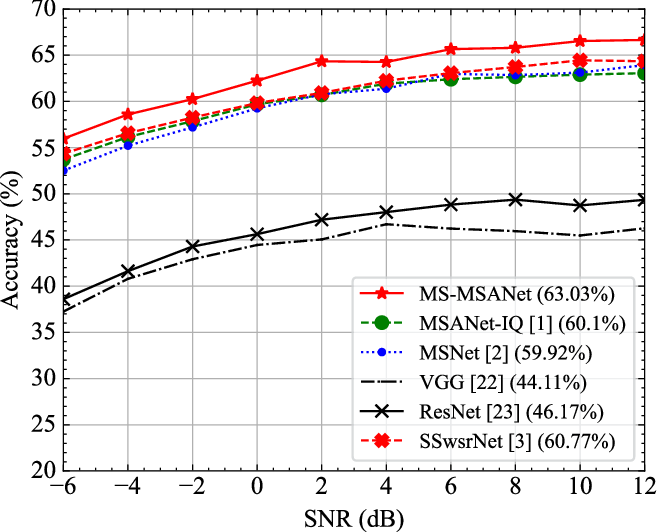} 
\caption{Classification performance comparison of the proposed FSOS-AMC framework under various signal-to-noise ratios (SNRs) using HisarMod2019.1 \cite{tekbiyik2020robust} under multipath fading channels and \emph{few-shot open-set} scenarios.}
\label{fig:fsos_2019}
\end{figure}

To further investigate the performance of the proposed FSOS-AMC framework under \emph{few-shot open-set} scenarios, we compare the classification performance of the proposed FSOS-AMC framework with the state-of-the-art methods, including VGG \cite{o2016convolutional}, ResNet \cite{o2018over}, MSNet \cite{zhang2021novel}, and SSwsrNet \cite{zhang2024sswsrnet} under various signal-to-noise ratio (SNR) levels using HisarMod2019.1 \cite{tekbiyik2020robust} under multipath fading channels, as shown in Fig. \ref{fig:fsos_2019}. 
The proposed FSOS-AMC framework, represented by MS-MSANet, demonstrates superior performance compared to state-of-the-art methods, achieving the highest accuracy of 76.72\%. It is followed closely by MSANet-IQ at 73.84\%. All models show improved performance as SNR increases, indicating better classification in less noisy conditions. The experiment, conducted using HisarMod2019.1 under multipath fading channels, provides a realistic scenario for evaluating these models in challenging signal environments. ResNet shows the lowest overall performance among the compared models, while the FSOS-AMC framework demonstrates its effectiveness in handling various noise levels in signal classification tasks.

\begin{figure}[!h]
\centering
\subfigure[]{
\includegraphics[width=0.95\linewidth]{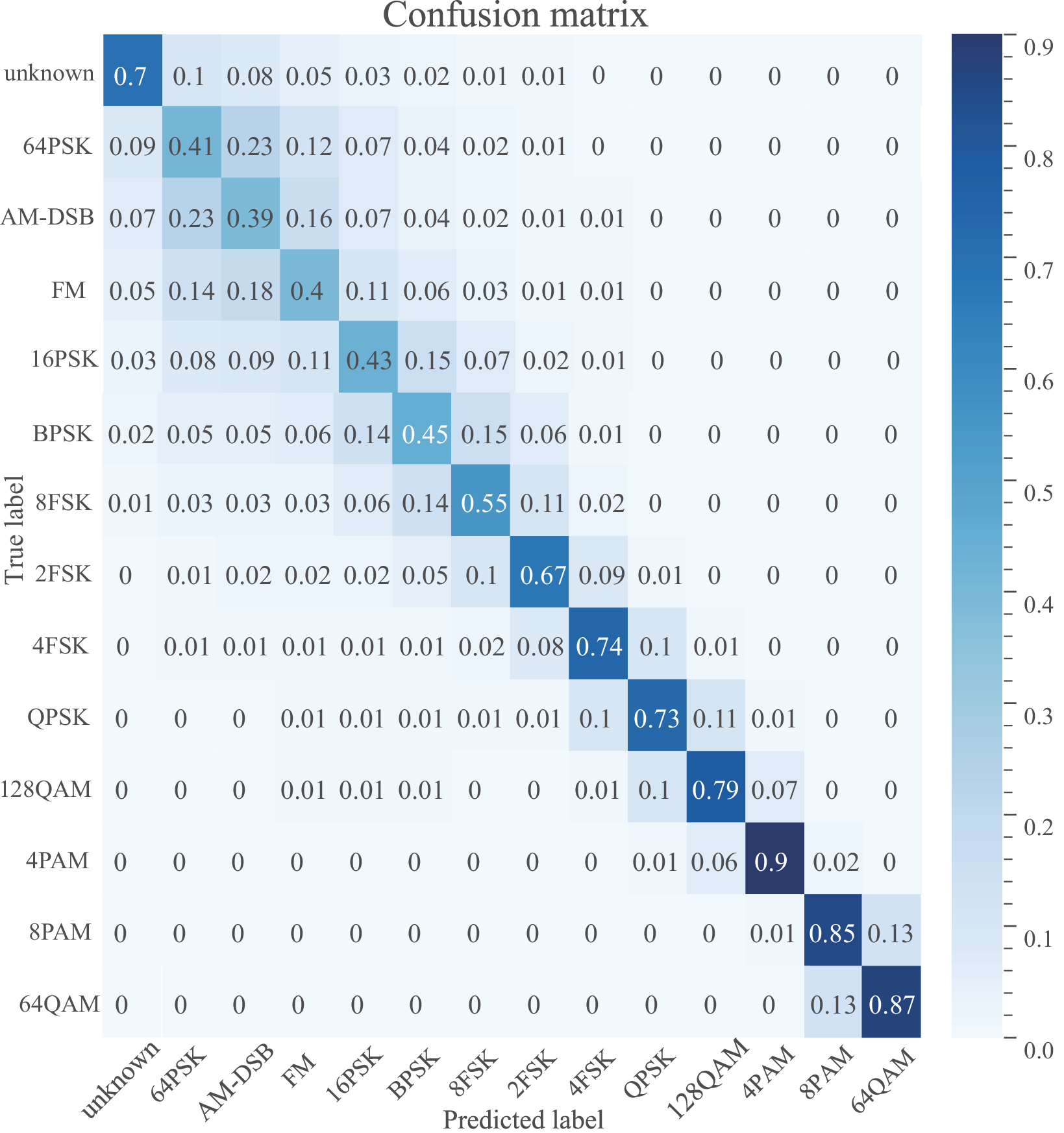}}
\subfigure[]{
\includegraphics[width=0.95\linewidth]{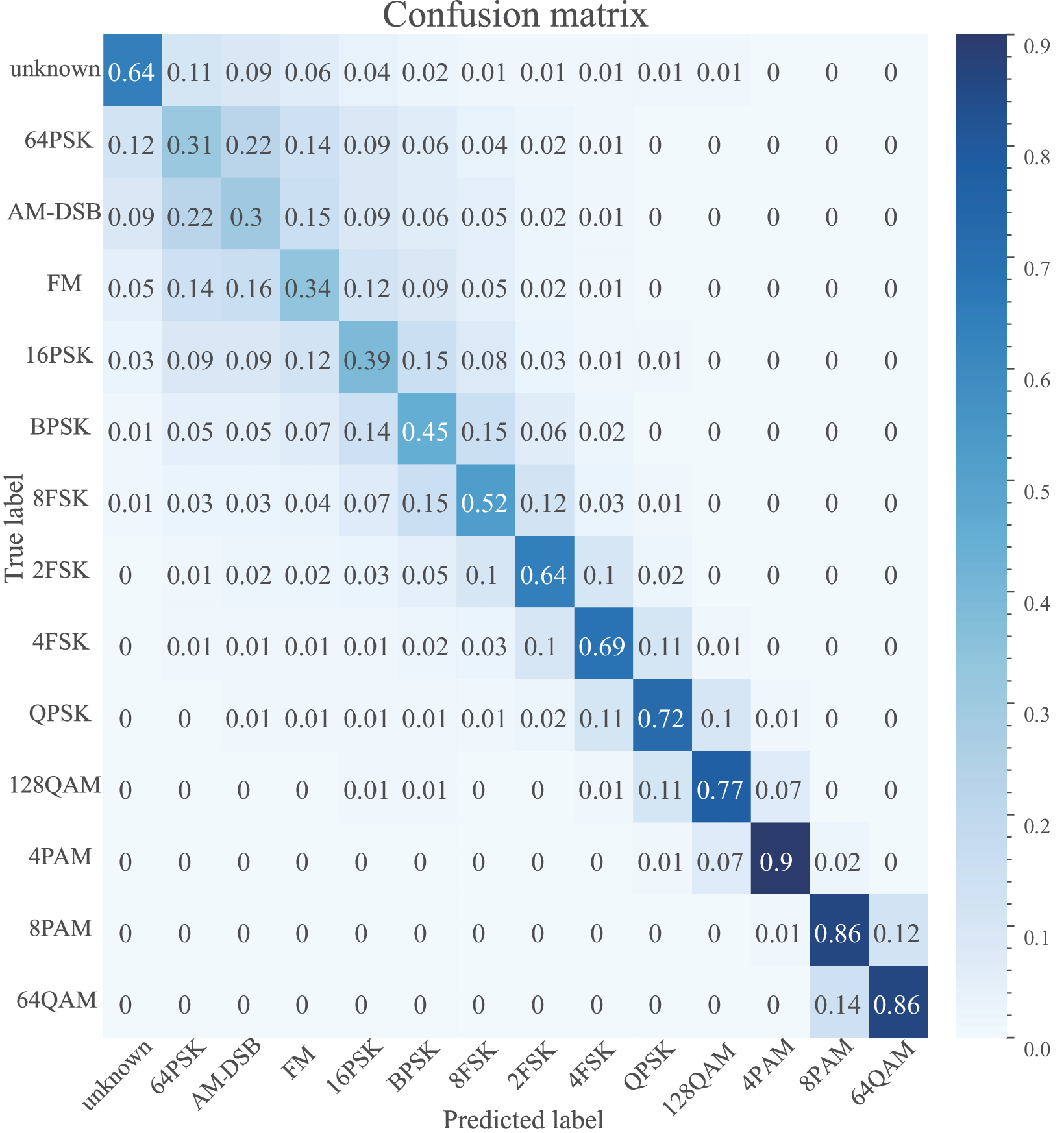}}
\DeclareGraphicsExtensions.
\caption{Confusion matrix of the proposed (a) FSOS-AMC framework (MS-MSANet) and (b) MSANet-IQ using HisarMod2019.1 \cite{tekbiyik2020robust} under multipath fading channels.}
\label{fig:confusion_matrix}
\end{figure}

For a more comprehensive evaluation of the proposed FSOS-AMC framework under \emph{few-shot open-set} scenarios, we present the confusion matrix of the proposed FSOS-AMC framework with different signal representations under multipath fading channels using HisarMod2019.1 \cite{tekbiyik2020robust}, as shown in Fig. \ref{fig:confusion_matrix}. 
The figure shows the classification performance across unknown modulation classes and known modulation classes, total 14 modulation classes. 
It can be seen that the proposed FSOS-AMC framework using multi-sequence signal representation achieves a better classification accuracy compared to the single IQ signal representation across all modulation classes except 8PAM. 
Moreover, for unknown modulation schemes, the proposed FSOS-AMC framework achieves a higher classification accuracy compared to MSANet-IQ with over $70\%$ accuracy (0.7 vs 0.64). 
The proposed FSOS-AMC framework demonstrates also shows better discrimination between similar modulations, evident in the sharper contrast between diagonal and off-diagonal elements. 
Notably, both models perform exceptionally well for higher-order modulations such as QAM and PAM, shown in the lower right corner with dark blue squares and minimal off-diagonal confusion. 

\begin{figure}
\centering
\includegraphics[width=0.99\linewidth]{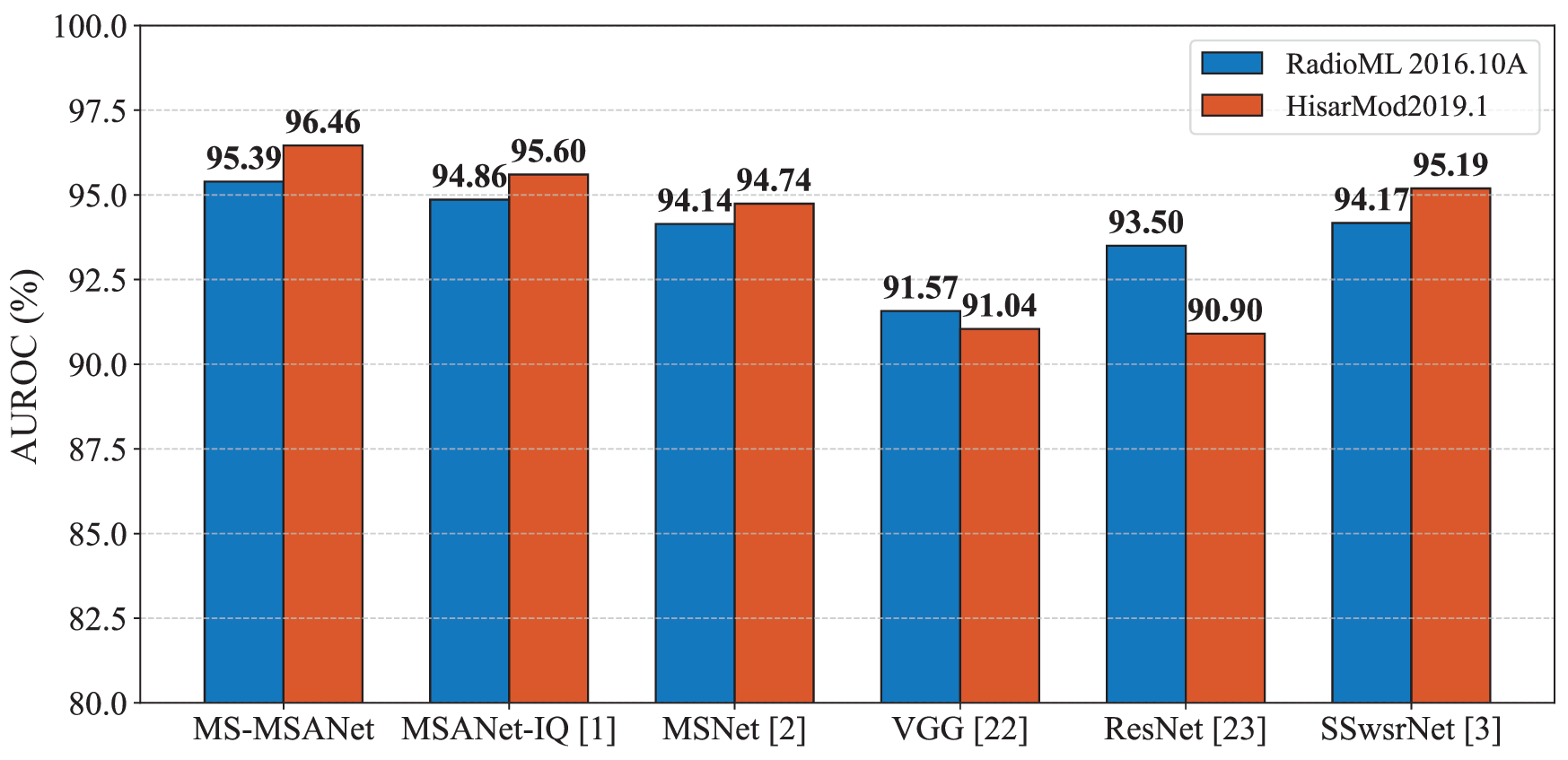} 
\caption{Performance comparison in terms of AUROC using RadioML2016.10a \cite{o2016convolutional} and HisarMod2019.1 \cite{tekbiyik2020robust}.}
\label{fig:auroc}
\end{figure}

To evaluate the performance of different models for open set recognition, we adopt AUROC for both RadioML 2016.10a \cite{o2016convolutional} and HisarMod2019.1 \cite{tekbiyik2020robust}, as shown in Fig. \ref{fig:auroc}. 
It can be observed that the proposed FSOS-AMC framework, represented by MS-MSANet, achieves the highest AUROC scores for both datasets, with 95.39\% for RadioML 2016.10A and 96.46\% for HisarMod2019.1. The MSANet-IQ \cite{zhang2024few} shows slightly lower performance, with AUROC scores of 94.86\% and 95.39\% for the two datasets, respectively. The other models, including VGG \cite{o2016convolutional}, ResNet \cite{o2018over}, MSNet \cite{zhang2021novel}, and SSwsrNet \cite{zhang2024sswsrnet}, exhibit lower AUROC scores, with VGG showing the lowest performance. The proposed FSOS-AMC framework demonstrates superior performance in open set recognition tasks, highlighting its effectiveness in classifying unknown modulation schemes under challenging conditions.

\section{Conclusion}
\label{sec:conclusion}

A novel few-shot open-set automatic modulation classification (FSOS-AMC) framework was proposed in this paper. 
The FSOS-AMC framework consists of three main parts, a multi-sequence multi-scale attention feature extractor (MS-MSANet), a meta-prototype training strategy, and a modular open-set classifier. 
The MS-MSANet was exploited to extract the features from the multi-sequence input signal, and the meta-prototype training strategy was utilized to train the feature extractor using a supervised, episode-based methodology. 
The modular open-set classifier was adopted to classify the testing data into one of the pre-defined known modulations or potential unknown modulations. 
Simulation results demonstrated that the proposed FSOS-AMC framework can achieve a high classification accuracy for both known modulations and unknown modulations. 
The proposed FSOS-AMC framework outperforms the recent advanced methods in terms of accuracy, and AUROC under few-shot open-set scenarios.

Despite the promising results of our FSOS-AMC framework, several challenges and opportunities remain for future research. The current framework faces challenges in balancing computational efficiency with classification accuracy for real-time applications, improving performance under extreme channel conditions, and handling dynamic spectrum environments. Future research directions include integrating transformer architectures to enhance feature extraction while reducing computational complexity, exploring meta-learning approaches to handle new modulation schemes with fewer training samples, and developing online learning capabilities for continuous adaptation to changing wireless environments. These challenges and opportunities suggest a rich landscape for future research in automatic modulation classification, particularly as wireless communications continue to evolve with new standards and requirements.

\bibliographystyle{IEEEtran}
\bibliography{open_set_ref}


\begin{IEEEbiography}[{\includegraphics[width=1in,height=1.25in,clip,keepaspectratio]{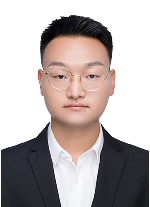}}]{Hao Zhang} (Graduate Student Member, IEEE) received the master's degree from the School of Information Engineering, Nanchang University, China, in 2020. He is currently pursuing the Ph.D. degree in the College of Electronic and Information Engineering, Nanjing University of Aeronautics and Astronautics, Nanjing, China. He was a visiting Ph.D. student at the School of Electrical \& Electronic Engineering, Nanyang Technological University, Singapore, in 2024. His research interests focus on deep learning, foundation models, wireless communication, signal processing, and spectrum cognition.\end{IEEEbiography}

\begin{IEEEbiography}[{\includegraphics[width=1in,height=1.25in,clip,keepaspectratio]{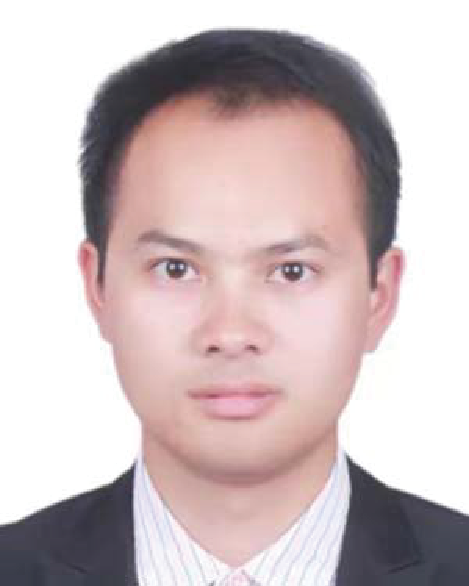}}]{Fuhui Zhou}
(Senior member, IEEE) is currently a Full Professor with Nanjing University of Aeronautics and Astronautics, Nanjing, China, where he is also with the Key Laboratory of Dynamic Cognitive System of Electromagnetic Spectrum Space. His research interests include cognitive radio, cognitive intelligence, knowledge graph, edge computing, and resource allocation. 

Prof. Zhou has published over 200 papers in internationally renowned journals and conferences in the field of communications. He has been selected for 1 ESI hot paper and 13 ESI highly cited papers. He has received 4 Best Paper Awards at international conferences such as IEEE Globecom and IEEE ICC. He was awarded as 2021 Most Cited Chinese Researchers by Elsevier, Stanford World’s Top 2\% Scientists, IEEE ComSoc Asia-Pacific Outstanding Young Researcher and Young Elite Scientist Award of China and URSI GASS Young Scientist. He serves as an Editor of IEEE Transactions on communication, IEEE Systems Journal, IEEE Wireless Communication Letters, IEEE Access and Physical Communications.
\end{IEEEbiography}
		
\begin{IEEEbiography}[{\includegraphics[width=1in,height=1.25in,clip,keepaspectratio]{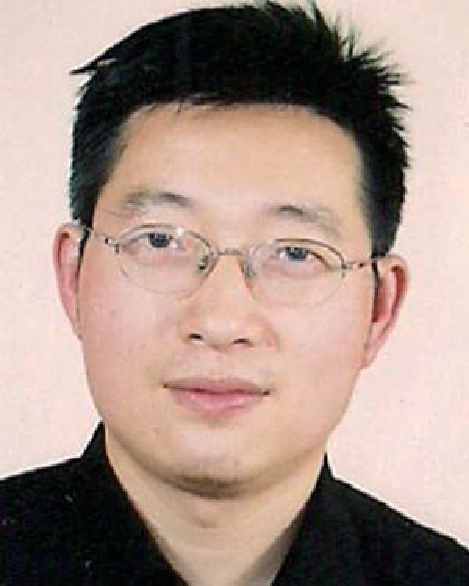}}]{Qihui Wu}
(Fellow, IEEE) received the B.S. degree in communications engineering, the M.S. and Ph.D. degrees in communications and information systems from the Institute of Communications Engineering, Nanjing, China, in 1994, 1997, and 2000, respectively. From 2003 to 2005, he was a Postdoctoral Research Associate with Southeast University, Nanjing, China. From 2005 to 2007, he was an Associate Professor with the College of Communications Engineering, PLA University of Science and Technology, Nanjing, China, where he was a Full Professor from 2008 to 2016. Since May 2016, he has been a Full Professor with the College of Electronic and Information Engineering, Nanjing University of Aeronautics and Astronautics, Nanjing, China. From March 2011 to September 2011, he was an Advanced Visiting Scholar with the Stevens Institute of Technology, Hoboken, USA. His current research interests span the areas of wireless communications and statistical signal processing, with emphasis on system design of software defined radio, cognitive radio, and smart radio.
\end{IEEEbiography}

\begin{IEEEbiography}[{\includegraphics[width=1in,height=1.25in,clip,keepaspectratio]{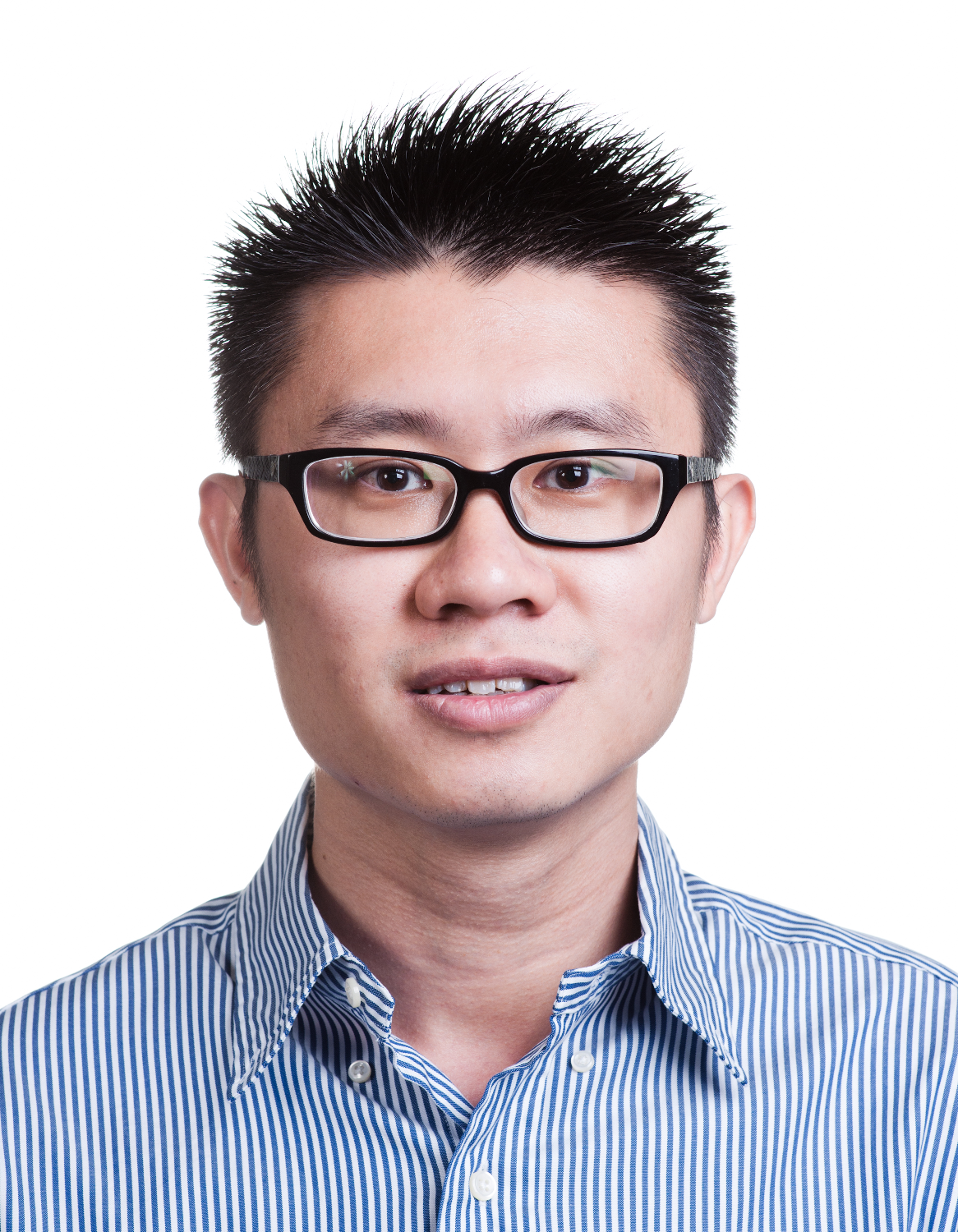}}]{Chau Yuen}
(S'02-M'06-SM'12-F'21) received the B.Eng. and Ph.D. degrees from Nanyang Technological University, Singapore, in 2000 and 2004, respectively. He was a Post-Doctoral Fellow with Lucent Technologies Bell Labs, Murray Hill, in 2005. From 2006 to 2010, he was with the Institute for Infocomm Research, Singapore. From 2010 to 2023, he was with the Engineering Product Development Pillar, Singapore University of Technology and Design. Since 2023, he has been with the School of Electrical and Electronic Engineering, Nanyang Technological University, currently he is Provost’s Chair in Wireless Communications, Assistant Dean in Graduate College, and Cluster Director for Sustainable Built Environment at ER@IN.

Dr. Yuen received IEEE Communications Society Leonard G. Abraham Prize (2024), IEEE Communications Society Best Tutorial Paper Award (2024), IEEE Communications Society Fred W. Ellersick Prize (2023), IEEE Marconi Prize Paper Award in Wireless Communications (2021), IEEE APB Outstanding Paper Award (2023), and EURASIP Best Paper Award for JOURNAL ON WIRELESS COMMUNICATIONS AND NETWORKING (2021).
 
Dr. Yuen current serves as an Editor-in-Chief for Springer Nature Computer Science, Editor for IEEE TRANSACTIONS ON VEHICULAR TECHNOLOGY, IEEE TRANSACTIONS ON NEURAL NETWORKS AND LEARNING SYSTEMS, and IEEE TRANSACTIONS ON NETWORK SCIENCE AND ENGINEERING, where he was awarded as IEEE TNSE Excellent Editor Award 2024 and 2022, and Top Associate Editor for TVT from 2009 to 2015. He also served as the guest editor for several special issues, including IEEE JOURNAL ON SELECTED AREAS IN COMMUNICATIONS, IEEE WIRELESS COMMUNICATIONS MAGAZINE, IEEE COMMUNICATIONS MAGAZINE, IEEE VEHICULAR TECHNOLOGY MAGAZINE, IEEE TRANSACTIONS ON COGNITIVE COMMUNICATIONS AND NETWORKING, and ELSEVIER APPLIED ENERGY.
 
He is a Distinguished Lecturer of IEEE Vehicular Technology Society, Top 2\% Scientists by Stanford University, and also a Highly Cited Researcher by Clarivate Web of Science. He has 4 US patents and published over 400 research papers at international journals.
\end{IEEEbiography}

\end{document}